\documentclass[11pt,letterpaper]{article}

\pdfoutput=1
\usepackage{jcappub}

\usepackage{appendix}
\usepackage{array}
\newcolumntype{x}[1]{>{\centering\arraybackslash}p{#1}}
\usepackage{bm}
\usepackage{color}
\usepackage{graphicx}
\usepackage{cancel}
\usepackage{amsfonts}
\usepackage{amssymb}
\usepackage{amsmath}
\usepackage{calc}
\usepackage{dcolumn}
\usepackage{mathrsfs}
\usepackage{mathtools}
\usepackage{soul}

\usepackage[normalem]{ulem} 

\newcommand{\eg}{e.g.~}
\newcommand{\ie}{i.e.~}

\newcommand{\Eq}[1]{Eq.~\eqref{#1}}
\newcommand{\Fig}[1]{Fig.~\ref{#1}}
\newcommand{\Sec}[1]{Sec.~\ref{#1}}
\newcommand{\App}[1]{Appendix \ref{#1}}

\newcommand{\Ord}{{\cal O}}		
\newcommand{\Hcur}{\mathcal{H}}

\newcommand{\GeV}{\,{\rm GeV}}

\newcommand{\beq}{\begin{equation}}
\newcommand{\eeq}{\end{equation}}

\newcommand{\ud}{\text{d}}
\newcommand{\bol}[1]{\boldsymbol{#1}}

\newcommand{\ER}{E_\text{R}}
\newcommand{\Ed}{E'}

\newcommand{\vmin}{v_\text{min}}

\newcommand{\teta}{\tilde{\eta}}

\newcommand{\ignore}[1]{}

\definecolor{rossoCP3}{cmyk}{0,.88,.77,.40}
\definecolor{verdeCP3}{rgb}{0.09765625, 0.57421875, 0.1015625}
\definecolor{bluCP3}{rgb}{0, 0.23, 0.67}

\ifx\pdfoutput\undefined
\usepackage[dvips,bookmarks]{hyperref}    
\else
\usepackage{hyperref}    
\fi
\hypersetup{colorlinks, bookmarksopen, bookmarksnumbered,
citecolor=verdeCP3, linkcolor=bluCP3, pdfstartview=FitH, urlcolor=rossoCP3}

\notoc

\newcommand{\AddrUCLA}{Department of Physics and Astronomy, UCLA, 475 Portola Plaza, Los Angeles, CA 90095 (USA)}

\newcommand{\AddrCERN}{CERN Theory Division, CH-1211, Geneva 23, Switzerland}

\begin{document}

\title{Assessing Compatibility of Direct Detection Data: Halo-Independent Global Likelihood Analyses}

\subheader{CERN-TH-2016-150}

\author[a]{Graciela B.~Gelmini,}
\author[b]{Ji-Haeng Huh,}
\author[a]{Samuel J.~Witte}

\affiliation[a]{\AddrUCLA}
\affiliation[b]{\AddrCERN}

\emailAdd{gelmini@physics.ucla.edu}
\emailAdd{ji-haeng.huh@cern.ch}
\emailAdd{switte@physics.ucla.edu}

\abstract{
We present two different halo-independent methods  to assess the compatibility of several  direct dark matter detection data sets for a given dark matter model using a global likelihood consisting of at least one extended likelihood and an arbitrary number of Gaussian or Poisson likelihoods. In the first method we find the global best fit halo function (we prove that it is a unique piecewise constant function with a number of down steps smaller than or equal to a maximum number that we compute) and construct a two-sided pointwise confidence band at any desired confidence level, which can then be compared with those derived from the extended likelihood alone to assess the joint compatibility of the data. In the second method we define a ``constrained parameter goodness-of-fit'' test statistic, whose $p$-value we then use to define a ``plausibility region'' (\eg where $p \geq 10\%$). For any halo function not entirely contained within the plausibility region, the level of compatibility of the data is very low (\eg $p < 10 \%$). We illustrate these methods by applying them to CDMS-II-Si and SuperCDMS data, assuming dark matter particles with elastic spin-independent isospin-conserving interactions or exothermic spin-independent isospin-violating interactions.}


\maketitle

\flushbottom

\newpage


\section{Introduction}
Astrophysical and cosmological evidence indicate that roughly $85 \%$ of the matter in the Universe is in the form of dark matter (DM) most likely composed of yet unknown elementary particles. Arguably the most extensively studied DM particle candidate is a weakly interacting massive particle (WIMP), which offers both theoretical appeal and hope for near-future detection. Most of the matter in our own galaxy resides in a spheroidal dark halo that extends much beyond the visible disk. Direct DM detection experiments represent one of the primary WIMP search methods currently employed. These experiments attempt to measure the recoil energy of nuclei after they collide with DM particles bound to the galactic dark halo passing through Earth. The current status of DM direct detection experiments remain ambiguous, with three experiments observing a potential DM signal and all others reporting upper bounds, some of which appear to be in irreconcilable conflict with the putative detection claims for most particle candidates~\cite{Bernabei:2010mq, Aalseth:2010vx, Aalseth:2012if, Aalseth:2011wp, Aalseth:2014eft, Aalseth:2014jpa, Agnese:2013rvf, Angle:2011th, Aprile:2011hi, Aprile:2012nq, Felizardo:2011uw, Archambault:2012pm, Behnke:2012ys, Ahmed:2012vq, Agnese:2015ywx,Akerib:2015rjg,Agnese:2015nto,Agnese:2014aze}. 

Interpreting the results of DM direct detection experiments typically requires assumptions on the local DM density, the DM velocity distribution, the DM-nuclei interaction, and the scattering kinematics. The uncertainties associated with these inputs can significantly affect the expected recoil spectrum (both in shape and magnitude) for a particular experiment, as well as the observed compatibility between experimental data. Attempts have been made to remove the astrophysical uncertainty from direct DM  detection calculations, and compare data in a ``halo-independent" manner, by translating measurements and bounds on the scattering rate into measurements and bounds on a function we will refer to as $\teta(\vmin,t)$ common to all experiments, which contains all of the information on the local DM density and velocity distribution (see \eg~\cite{Fox:2010bz,Fox:2010bu,Frandsen:2011gi,Gondolo:2012rs,HerreroGarcia:2012fu,Frandsen:2013cna,DelNobile:2013cta,Bozorgnia:2013hsa,DelNobile:2013cva,DelNobile:2013gba,DelNobile:2014eta,Feldstein:2014gza,Fox:2014kua,Gelmini:2014psa,Cherry:2014wia,DelNobile:2014sja,Scopel:2014kba,Feldstein:2014ufa,Bozorgnia:2014gsa,Blennow:2015oea,DelNobile:2015lxa,Anderson:2015xaa,Blennow:2015gta,Scopel:2015baa,Ferrer:2015bta,Wild:2016myz,Kahlhoefer:2016eds}).  

The function $\teta(\vmin,t)$ depends on the time $t$ and a particular speed $\vmin$. The physical interpretation of $\vmin$ depends on the type of analysis being used. If the nuclear recoil $\ER$ is considered an independent variable, then $\vmin$ is understood to be the minimum speed necessary for the incoming DM particle to impart a nuclear recoil $\ER$ to the target nucleus (and thus it depends on the target nuclide $T$ through its mass $m_T$, $\vmin^T=\vmin(\ER,m_T)$). This has been the more common approach~\cite{Fox:2010bz,Frandsen:2011gi,Frandsen:2013cna}. Alternatively, one can choose $\vmin$ as the independent variable, in which case $\ER^{T}$ is understood to be the extremum recoil energy (maximum for elastic scattering, and either maximum or minimum for inelastic scattering) that can be imparted by an incoming WIMP traveling with speed $v = \vmin$ to a target nuclide $T$. Note that for elastic scattering off a single nuclide target the two approaches are just related by a simple change of variables. We will choose to treat $\vmin$ as an independent variable for the remainder of this paper, as this choice allows us to account for any isotopic target composition by summing terms dependent on $\ER^{T}(\vmin)$ over target nuclides $T$, for any fixed detected energy $\Ed$.

Early halo-independent analyses were limited in the way they handled putative signals. Only weighted averages on $\vmin$ intervals of the unmodulated component of $\teta(\vmin,t)$, $\teta^0(\vmin)$, and of the amplitude of the annually modulated component, $\teta^1(\vmin)$, (see \Eq{eta} below) were plotted against upper bounds in the $\vmin - \teta$ plane (see \eg \cite{Fox:2010bz,Frandsen:2011gi,Gondolo:2012rs,DelNobile:2013cva}). This type of analysis leads to a poor understanding of the compatibility of various data sets. 

Recently, attempts have been made to move beyond this limited approach of taking averages over $\vmin$ intervals by finding a best fit $\teta^0$ function and constructing confidence bands in the $\vmin - \teta$ plane \cite{Fox:2014kua,Gelmini:2015voa}, from unbinned data with an extended likelihood~\cite{Barlow:1990vc}. One can then compare upper bounds at a particular confidence level (CL) with a confidence band at a particular CL to assess if they are compatible (see \cite{Gelmini:2015voa} for a discussion). From now on, when an upper index $0$ or $1$ is not written, $\teta(\vmin)$ is understood to be $\teta^0(\vmin)$.

An alternative approach to analyzing the compatibility of data has been studied in~\cite{Feldstein:2014ufa} using the ``parameter goodness-of-fit'' test statistic \cite{Maltoni:2002xd,Maltoni:2003cu} derived from a global likelihood (an alternative approach is taken in \cite{Bozorgnia:2014gsa}). In \cite{Feldstein:2014ufa}, the compatibility of various experiments within a particular theoretical framework was determined by obtaining a $p$-value from Monte Carlo (MC) simulated data, generated under the assumption that the true halo function is the global best fit halo function. This approach has an advantage in that one can make quantitative statements about the compatibility between the observed data given a dark matter candidate model. However, this procedure assigns only a single number to the whole halo-independent parameter space, and we would like to have the ability to assess compatibility of the data with less restrictive assumptions on the underlying halo function.

In this paper we extend the approaches of~\cite{Feldstein:2014ufa} and~\cite{Gelmini:2015voa} by using the global likelihood function to assess the compatibility of multiple data sets within a particular theoretical model across the halo-independent $\vmin-\teta$ parameter space. This is done with two distinct approaches. First, we extend the construction of the halo-independent pointwise confidence band presented in~\cite{Gelmini:2015voa} to the case of a global likelihood function, consisting of one (or more) extended likelihood functions and an arbitrary number of Gaussian or Poisson likelihoods. The resultant global confidence band can be compared directly with the confidence band constructed from the extended likelihood alone, to assess the joint compatibility of the data for any choice of DM-nuclei interaction and scattering kinematics. The drawback of this method is that it cannot quantitatively address the level of compatibility of the data sets. To address this concern we also propose an extension of the parameter goodness-of-fit test, which we will refer to as the ``constrained parameter goodness-of-fit" test, that quantifies the compatibility of various data sets for a given DM particle candidate assuming the halo function $\teta(\vmin)$ passes through a particular point $(v^*,\teta^*)$. By calculating the $p$-values for each $(v^*,\teta^*)$ throughout the $\vmin-\teta$ plane, one can construct plausibility regions, such that for any halo function not entirely contained within the plausibility region the data are incompatible at the chosen level, e.g. $p < 10\%$.

In Sec.~\ref{haloindep} we review the procedure for constructing the best fit halo function $\teta_{BF}$ and confidence band from an extended likelihood. Readers familiar with~\cite{Gelmini:2015voa} may wish to skip this section and go directly to Sec.~\ref{sec:Extension}, which  discusses how the construction of the best fit halo function and confidence band is altered when dealing with a global likelihood function that is the product of one (or more) extended likelihoods and an arbitrary number of Poisson or Gaussian likelihoods. In Sec.~\ref{globalband1}, we use the methods discussed in Sec.~\ref{sec:Extension} to construct the best fit halo and global pointwise confidence band, for the combined analysis of CDMS-II-Si and SuperCDMS data assuming elastic isospin-conserving~\cite{Kurylov:2003ra,Chang:2010yk,Feng:2011vu} and exothermic isospin-violating spin-independent (SI) interactions~\cite{Gelmini:2014psa,Scopel:2014kba}. Sec.~\ref{sec:constrained} introduces the ``constrained parameter goodness-of-fit'' test statistic and the construction of the plausibility regions. This method is illustrated using CDMS-II-Si and SuperCDMS data, assuming elastic isospin-conserving spin-independent interactions. We conclude in Sec.~\ref{conclusion}.

\section{Review of the Extended Maximum-Likelihood Halo-independent (EHI) Analysis Method \label{haloindep}}
\subsection{Generalized halo-independent analysis}
The differential rate per unit of detector mass as a function of nuclear recoil energy $\ER$ for dark matter particles of mass $m$ scattering off a target nuclide $T$ with mass $m_T$ is given by
\begin{equation}
\label{diffrate}
\frac{\ud R_T}{\ud \ER} = \frac{\rho}{m}\frac{C_T}{m_T}\int_{v \geqslant \vmin(\ER)} \, \ud^3 \, v \, f(\bol{v},t) \, v \, \frac{\ud \sigma_T}{\ud \ER}(\ER, \bol{v}) \, ,
\end{equation}
where $\rho$ is the local dark matter density, $C_T$ is the mass fraction of the nuclide $T$ in the detector, $f(\bol{v},t)$ is the dark matter velocity distribution in Earth's frame, and $\ud \sigma_T / \ud \ER$ is the WIMP-nuclide differential cross section in the lab frame. When multiple target elements are present in the detector, the differential rate is
\begin{equation}\label{sum_diffrate}
\frac{\ud R}{\ud \ER} = \sum_T \frac{\ud R_T}{\ud \ER} \, .
\end{equation}

To allow for the possibility of inelastic DM-nuclei scattering, we consider a DM particle scattering to a new state of mass $m^{\prime} = m + \delta$, where $|\delta| \ll m$, and $\delta > 0$ ($<0$) describes endothermic (exothermic) scattering. In the limit $\mu_T |\delta|/m^2 \ll 1$, $\vmin(\ER)$ is given by 
\begin{equation}\label{eq:vmin}
\vmin(\ER) = \frac{1}{\sqrt{2 m_T \ER}} \left| \frac{m_T \ER}{\mu_T} +\delta \right| \, ,
\end{equation}
where $\mu_T$ is the reduced mass of the WIMP-nucleus system. Notice \Eq{eq:vmin} reduces to the typical equation for elastic scattering when $\delta = 0$. \Eq{eq:vmin} can be used to obtain the range of possible recoil energies, $[\ER^{T,-}(v),\ER^{T,+}(v)]$, that can be imparted to a target nucleus by a DM particle traveling at speed $v$ in Earth's frame, given by
\begin{equation}\label{eq:Ebranch}
\ER^{T,\pm} (v) = \frac{\mu_T^2 v^2}{2 m_T} \left( 1 \pm \sqrt{1-\frac{2\delta}{\mu_T v^2}} \right)^2 \, .
\end{equation}
\Eq{eq:Ebranch} shows that for endothermic scattering there exists a nontrivial kinematic endpoint, given by the DM speed $v_\delta^T = \sqrt{2 \delta / \mu_T}$, below which incoming DM particles cannot induce nuclear recoils. When multiple targets are present in a detector, we use $v_\delta$ to denote the minimum of all $v_\delta^T$. For exothermic and elastic scattering $v_\delta=0$.

Experiments do not actually measure the recoil energy of a target nucleus, but rather a proxy for recoil energy (\eg the number of photoelectrons detected in a photomultiplier tube) denoted $\Ed$. The differential rate as a function of the detected energy $\Ed$ is given by
\begin{equation}\label{eq:diffrate_ep}
\frac{\ud R}{\ud \Ed} = \sum_T \int_0^\infty \ud \ER \, \epsilon(\ER,\Ed) \, G_T(\ER,\Ed) \, \frac{\ud R_T}{\ud \ER} \, ,
\end{equation}
where the differential rate in \Eq{diffrate} has been convolved with the efficiency function $\epsilon(\ER,\Ed)$ and the energy resolution function $G_T(\ER,\Ed)$, which together give the probability that a detected recoil energy $\Ed$ resulted from a true recoil energy $\ER$.

Upon changing the order of integration, one can express the differential rate in detected energy as
\begin{equation}\label{diffrate_manip1}
\frac{\ud R}{\ud \Ed} = \frac{\sigma_\text{ref} \rho}{m} \int_{v \geqslant v_\delta} \ud^3 v \, \frac{f(\bol{v},t)}{v} \, \sum_T \frac{\ud \Hcur_T}{\ud \Ed} (\Ed, \bol{v}) \, ,
\end{equation}
where $\ud \Hcur_T/\ud \Ed$ is given by
\begin{equation}\label{eq:dHcurl}
 \frac{\ud \Hcur_T}{\ud \Ed}(\Ed, \bol{v}) \equiv 
  \begin{dcases} 
      \hfill \frac{C_T}{m_T}\int_{\ER^{T,-}}^{\ER^{T,+}} \, \ud \ER \epsilon(\ER,\Ed) \, G_T(\ER, \Ed) \, \frac{v^2}{\sigma_\text{ref}} \, \frac{\ud \sigma_T}{\ud \ER}(\ER, \bol{v})    \hfill & \text{ if $v \geqslant v_\delta^T$,} \\
      \hfill 0 \hfill & \text{ if $v < v_\delta^T$.} \\
  \end{dcases}
\end{equation}
and we define 
\begin{equation}
\frac{\ud \Hcur}{\ud \Ed} \equiv \sum_T  \frac{\ud \Hcur_T}{ \ud \Ed} \, .
\end{equation}
Here, we only consider differential cross sections that depend on the speed of the WIMP $v = |\bol{v}|$. The cross section depends only on the speed $v$ if the incoming WIMPs and the target nuclei are unpolarized and the detector response is isotropic, as is most common. In Eqs.~\ref{diffrate_manip1} and~\ref{eq:dHcurl}, we have incorporated the parameter $\sigma_\text{ref}$ which denotes the overall strength of the interaction. For example in the case of the SI interaction, with differential cross section given by
\begin{equation}
\frac{\ud \sigma_T^{SI}}{\ud \ER}(\ER,v) = \sigma_p \frac{\mu_T^2}{\mu_p^2}[Z_T+(A_T-Z_T)(f_n/f_p)]^2 \, \frac{F_T^2(\ER)}{2 \mu_T^2 v^2 / m_T} \, ,
\end{equation}
where $A_T$ and $Z_T$ are the atomic and charge numbers of nuclide $T$, $f_n$ and $f_p$ are the neutron and proton couplings, and $F_T(\ER)$ is the form factor normalized to $F_T(0)=1$ (taken here to be Helm form factor), we will choose $\sigma_\text{ref} = \sigma_p$, the WIMP-proton cross section.

A halo-independent analysis relies on the separation of the astrophysical parameters from the particle physics and detector-dependent quantities. Here we follow \cite{DelNobile:2013cva}. Let us define
\begin{equation}\label{eq:eta_t}
\tilde{\eta}(\vmin,t) \equiv \frac{\rho \sigma_\text{ref}}{m}\int_{\vmin}^{\infty} \, \ud v \, \frac{F(v,t)}{v} \, ,
\end{equation}
where $F(v,t) \equiv v^2 \int \ud \Omega_v f(\bol{v},t)$. Differentiating both sides of \Eq{eq:eta_t} gives 
\begin{equation}
\frac{\sigma_\text{ref}\rho}{m} \, \frac{F(v,t)}{v} = - \frac{\partial \tilde{\eta}(v,t)}{\partial v} \, ,
\end{equation}
which upon insertion into \Eq{diffrate_manip1} leads to
\begin{equation}\label{drate_detatilde}
\frac{\ud R}{\ud \Ed} = - \int_{v_\delta}^{\infty} \ud v \, \frac{\partial \tilde{\eta}(v,t)}{\partial v} \, \frac{\ud \Hcur}{\ud \Ed}(\Ed, v) \, .
\end{equation}
Using the fact that $\tilde{\eta}(\infty,t) = 0$ and $\ud \Hcur / \ud \Ed (\Ed, v_\delta) = 0$, integration by parts of \Eq{drate_detatilde} results in the following expression for the differential rate
\begin{equation}
\frac{\ud R}{\ud \Ed} = \int_{v_\delta}^{\infty} \ud \vmin \tilde{\eta}(\vmin, t) \, \frac{\ud \mathcal{R}}{\ud \Ed}(\Ed, \vmin) \, ,
\end{equation}
where we have now defined the differential response function $\ud \mathcal{R}/ \ud \Ed$ as 
\begin{equation}
\frac{\ud \mathcal{R}}{\ud \Ed}(\Ed,\vmin) \equiv \frac{\partial}{\partial \vmin}\left[ \frac{\ud \Hcur}{\ud \Ed}(\Ed,\vmin) \right] \, .
\end{equation}

$\teta(\vmin,t)$ is a function of time due to the annual rotation of the Earth around the Sun. If one now makes the approximation 
\begin{equation} \label{eta}
\tilde{\eta}(\vmin,t) \simeq \tilde{\eta}^0(\vmin) + \tilde{\eta}^1(\vmin) \cos(2 \pi (t-t_0)/\text{year})
\end{equation}
and integrates the differential rate over the energy range of interest, the unmodulated component $R^0$ and annual modulation amplitude $R^1$ of the rate are given by
\begin{align}\label{eq:rateeq}
R^{\alpha}_{[\Ed_1,\Ed_2]} & \equiv \int_{v_\delta}^{\infty} \ud \vmin \, \tilde{\eta}^\alpha (\vmin) \, \int_{\Ed_1}^{\Ed_2} \, \ud \Ed \, \frac{\ud \mathcal{R}}{\ud \Ed}  \\ &
 = \int_{v_\delta}^{\infty} \ud \vmin \, \tilde{\eta}^\alpha (\vmin) \, \mathcal{R}_{[\Ed_1,\Ed_2]}(\vmin)  \, ,
\end{align}
where $\alpha = 0$ or $1$, and the energy-integrated response function $\mathcal{R}$ is given by
\begin{equation}
\mathcal{R}_{[\Ed_1,\Ed_2]}(\vmin) = \int_{\Ed_1}^{\Ed_2} \, \ud \Ed \, \frac{\ud \mathcal{R}}{\ud \Ed}(\Ed, \vmin) \, .
\end{equation}
In the event that $\mathcal{R}_{[\Ed_1,\Ed_2]}(\vmin)$ is a well-localized function in $\vmin$, measurements on unmodulated and modulated rate can be used to infer the average values of $\tilde{\eta}^0$ and $\tilde{\eta}^1$ over a $\vmin$ interval. This is the case for DM-nuclei differential cross sections proportional to $1/v^2$ (\eg the typical SI and SD contact interactions). Should the differential cross section not be of this form, one may need to regularize the energy-integrated response function as described in \cite{DelNobile:2013cva}.

\subsection{Extended maximum likelihood analysis \label{ehianalysis}}
It was initially proven in \cite{Fox:2014kua}, that if there is no uncertainty in the measurement of recoil energies in a single nuclide target, then the extended likelihood, given by
\begin{equation}\label{eq:extendedlikelihood}
\mathcal{L}[\tilde{\eta}(\vmin)] \equiv e^{-N_E[\tilde{\eta}]} \prod_{a = 1}^{N_O}MT \, \frac{\ud R_\text{tot}}{\ud \Ed} \biggr{\rvert}_{\Ed = \Ed_a} \, ,
\end{equation}
is maximized by a non-increasing piecewise constant $\tilde{\eta}^0(\vmin)$ function (which we call simply $\teta(\vmin)$) with at most $N_O$ (the number of observed events) steps. $N_E[\teta]$ in \Eq{eq:extendedlikelihood} is the total number of expected events, and $\Ed_a$ is the observed energy of event $a$. This proof was generalized to the case of realistic energy resolution and arbitrary target composition in \cite{Gelmini:2015voa}. The generalized proof presented in \cite{Gelmini:2015voa} applies the Karush-Kuhn-Tucker (KKT) conditions, which are only valid for systems with an objective function of finite number of variables subject to a finite number of constraints, to the likelihood functional in \Eq{eq:extendedlikelihood} by discretizing the variable $\vmin$, applying the KKT conditions, and then taking the continuum limit. 

Here, we will briefly review the conclusions presented in \cite{Gelmini:2015voa}. If one defines the quantity
\begin{equation}\label{loglike}
L[\tilde{\eta}] = -2 \ln{\mathcal{L}[\tilde{\eta}]} \, ,
\end{equation}
then instead of maximizing the likelihood, one can equivalently minimize $L[\tilde{\eta}]$. The KKT conditions, applied to \Eq{loglike} and taken in the continuum limit, lead to the following:
\begin{flalign}
\text{(I)} &  \hspace{.5cm} q(\vmin) = \int_{v_\delta}^{\vmin}\ud v \, \frac{\delta L}{\delta \tilde{\eta}(v)} \label{eq:qvmin1} \\ \vspace{.2cm}
\text{(II)} & \hspace{.5cm} q(\vmin) \geq 0 \label{eq:qvmin2} \\ 
\text{(III)} & \hspace{.5cm} \forall \epsilon > 0, \hspace{.5cm} \tilde{\eta}(\vmin+\epsilon) \leq \tilde{\eta}(\vmin) \label{eq:qvmin3} \\ \vspace{.2cm} 
\text{(IV)} & \hspace{.5cm} q(\vmin) \lim_{\epsilon \rightarrow +0} \frac{\tilde{\eta}(\vmin + \epsilon) - \tilde{\eta}(\vmin)}{\epsilon} = 0 \label{eq:qvmin4} \, .
\end{flalign}
A direct consequence of \Eq{eq:qvmin4} is that $\teta(\vmin)$ is a piecewise constant function with the locations of the steps given by the $\vmin$ values which satisfy $q(\vmin) = 0$. For this reason, we need to analyze the behavior of $q(\vmin)$. \Eq{eq:extendedlikelihood} and~(\ref{eq:qvmin1}) can be used to show that
\begin{equation}\label{extendedlilq}
q(\vmin) = 2 \xi(\vmin) - 2 \sum_{a = 1}^{N_O} \frac{H_a(\vmin)}{\gamma_a[\tilde{\eta}]} \, ,
\end{equation}
where we have defined the following quantities:
\begin{equation}\label{eq:xi}
\xi(\vmin) \equiv MT \int_{\Ed_\text{min}}^{\Ed_\text{max}} \ud \Ed \, \frac{\ud \Hcur}{\ud \Ed} (\Ed, \vmin) \, ,
\end{equation}
\begin{equation}
H_a(\vmin) \equiv \frac{\ud \Hcur}{\ud \Ed}(\Ed, \vmin) \biggr{\rvert}_{\Ed = \Ed_a} \, ,
\end{equation}
and
\begin{equation}
\gamma_a [\tilde{\eta}] \equiv \frac{\ud R_\text{tot}}{\ud \Ed} \biggr{\rvert}_{\Ed = \Ed_a} \, .
\end{equation}

For the extended likelihood function in \Eq{eq:extendedlikelihood}, the behavior of the terms in \Eq{extendedlilq} were studied in~\cite{Gelmini:2015voa} to determine how many steps can appear in the best fit $\teta$ function. We briefly review their behavior here (see \cite{Gelmini:2015voa} for additional details). 

Consider first the $\vmin$-dependence of $\ud \Hcur/ \ud \Ed$ (see \Eq{eq:dHcurl}), which appears in both the integrand of $\xi(\vmin)$ and in $H_a(\vmin)$. If the differential cross section is proportional to $v^{-2}$, as is the case for the standard SI and SD contact interactions, the only velocity dependence of $\ud \Hcur/ \ud \Ed$ is in the integration range $[E_R^{T,+}(\vmin),E_R^{T,-}(\vmin)]$. For these interactions, as $\vmin$ increases, the integration covers a larger portion of the parameter space where the integrand is non-zero. At large values of $\vmin$, the entire region where the integrand is non-zero is included in the integration and $\ud \Hcur/ \ud \Ed$ becomes constant. For a fixed value of $\Ed$, one would expect the integrand of $\ud \Hcur/ \ud \Ed$ to be a well-localized function of $\ER$ (\ie an observed recoil $\Ed$ can only result from a narrow range of $\ER$ values). For this reason, the terms $H_a(\vmin)$ appear as step-like functions in $\vmin$.

The term $\xi(\vmin)$ contains an additional integration of $\ud \Hcur/ \ud \Ed$ over $\Ed$. The only dependence on $\Ed$ appears in the factor $\epsilon(\Ed,\ER)G_T(\Ed,\ER)$, which describes the probability a detected recoil energy $\Ed$ is the result of some true recoil energy $\ER$. For small values of $\vmin$, only a narrow range of recoil energies are integrated over and thus $\xi$ will be quite small (\ie, for $\vmin$ values such that $\ER^{T,+}(\vmin)$ is below threshold). As $\vmin$ increases, the integration range widens and $\xi(\vmin)$ steadily increases. Eventually, the entire region where the integrand is nonzero is included in the integration, and $\xi(\vmin)$ becomes constant. 

The only term dependent on the halo function is $\gamma_a[\teta]$, which only alters the relative contribution of each step-like function to $q(\vmin)$. The function $\teta_{BF}(\vmin)$ can only be discontinuous when $q(\vmin) = 0$, which is equivalent to saying the steps of $\teta_{BF}$ occur where the step-like functions $H_a(\vmin)/\gamma_a[\teta]$ touch $\xi(\vmin)$ from below. Since there is a single term of the form $H_a(\vmin)/\gamma_a[\teta]$ for each observed event, the number of steps appearing in $\teta_{BF}$ must be less than or equal to the number of observed events, $N_O$.

\subsection{Construction of the best fit halo function and confidence band from an extended likelihood\label{sec:confidencebands}}
In this section we briefly review the construction of the best fit function $\teta_{BF}(\vmin)$ and the confidence band for an extended likelihood~\cite{Gelmini:2015voa}. Let us define the function $f_L^{N_O}$ of $2N_O$ variables, 
\begin{equation}
f_L^{N_O}(\vec{v},\vec{\teta}) \equiv L[\teta^{N_O}(\vmin;\vec{v},\vec{\teta})] \, ,
\end{equation}
where $\vec{v} = (v_1,...,v_{N_O})$ and $\vec{\teta}=(\teta_1,...,\teta_{N_O})$, and the various $v_a$ and $\teta_a$ specify the location and height of each step. Here, we have defined the piecewise constant function $\teta^{N_O}$ as 
\[
\teta^{N_O}(\vmin;\vec{v},\vec{\teta}) \equiv  
    \begin{cases} 
      \teta_a & v_{a-1} < \vmin \leq v_a  \, ,
      \\
      0 & v_{N_O} < \vmin \, .
   \end{cases} 
\]
Using the result of the previous section, minimizing the functional $L[\teta]$, and thus finding the best fit $\teta(\vmin)$, is now reduced to minimizing $f_L^{N_O}$ subject to the constraints 
\begin{equation}
v_1 > v_\delta \, ,
\end{equation}
\begin{equation}
v_b - v_a \geq 0 \, \text{ and } \, \teta_a - \teta_b \geq 0 \, \text{ for } \, a<b \, .
\end{equation}

We can define the confidence band as the region filled by all possible $\teta$ functions satisifying  
\begin{equation}\label{CBdef}
\Delta L[\teta] \equiv L[\teta] - L_\text{min} \leq \Delta L^* \, ,
\end{equation}
where $L_\text{min}$ is the minimum of $L[\teta]$, and $\Delta L^*$ corresponds to the desired confidence level. However, in practice, finding all $\teta$ functions satisfying \Eq{CBdef} is not possible. Instead, let us consider the possible subset of $\teta$ functions which minimize $L[\teta]$ subject to the constraint 
\begin{equation}\label{constrain}
\teta(v^*) = \teta^* \, .
\end{equation}
Now let us define $L_\text{min}^{c}(v^*,\teta^*)$ to be the minimum of $L[\teta]$ subject to the constraint in \Eq{constrain}, and 
\begin{equation}
\Delta L_\text{min}^{c}(v^*,\teta^*) = L_\text{min}^{c}(v^*,\teta^*) - L_\text{min} \, .
\end{equation}
If the point $(v^*,\teta^*)$ lies within the confidence band, then there should exist at least one $\teta$ function passing through this point which satisfies $\Delta L[\teta] \leq \Delta L^*$. Should this be the case, it follows that $\Delta L_\text{min}^{c}(v^*,\teta^*) \leq \Delta L^*$. Alternatively, if $\Delta L_\text{min}^{c}(v^*,\teta^*) \geq \Delta L^*$, one can state that there does not exist a single $\teta$ which satisfies $\Delta L[\teta] \leq \Delta L^*$. Thus the confidence band can be constructed by finding the values of $(v^*,\teta^*)$ which satisfy $\Delta L_\text{min}^{c}(v^*,\teta^*) \leq \Delta L^*$. This condition defines a two-sided interval around $\tilde\eta_{\rm BF}$ for each $\vmin$ value (with $\vmin=v^*$), and the collection of those intervals forms a pointwise confidence band in $\vmin$--$\tilde\eta$ space, which we are simply
calling the confidence band.

To understand the meaning of $\Delta L_\text{min}^c$, let us first discretize the continuous variable $\vmin$ into a collection of $K$ discrete values
 $\vec{v}_\text{min} = (\vmin^0,...,\vmin^{K-1})$. The likelihood functional in \Eq{loglike} then becomes a function of the $K-$dimensional vector $\vec{\teta} = (\teta_0,\teta_1,...,\teta_{K-1})$ which defines the piecewise constant function $\teta(\vmin;\vec{\teta})$ given by
\begin{equation}
\teta(\vmin;\vec{\teta}) \equiv \teta_i \text{  if  } \vmin^i \leq \vmin < \vmin^{i+1} \, .
\end{equation}
With this discretization, the constraint on $(v^*,\teta^*)$ in \Eq{constrain} corresponds to $\vmin^k \leq v^* < \vmin^{k+1}$ and $\teta^* = \teta_k$ for some integer $0 \leq k \leq K-1$. $\Delta L_\text{min}^{c}(v^*,\teta^*)$ is then replaced by the function $\Delta L_\text{min}^{k,c}(\teta^*)$ with the index $k$ corresponding to $v^*$, defined by
\begin{equation}\label{consproflike}
\Delta L_\text{min}^{k,c}(\teta^*) = - 2 \ln \left[\frac{\mathcal{L}(\hat{\hat{\teta}}_0,...,\hat{\hat{\teta}}_{k-1},\teta_k=\teta^*,\hat{\hat{\teta}}_{k+1},...,\hat{\hat{\teta}}_{K-1})}{\mathcal{L}(\hat{\teta}_0,...,\hat{\teta}_k,...,\hat{\teta}_{K-1})} \right] \, , 
\end{equation}
where $\hat{\hat{\teta}}_i$ are the $\teta_i$ values which maximize the likelihood function $\mathcal{L}(\teta_0,...\teta_{K-1}) \equiv \mathcal{L}[\teta(\vmin; \vec{\teta})]$ subject to the constraint $\teta_k = \teta^*$, and $\hat{\teta}_i$ maximize $\mathcal{L}$ without the constraint. $\Delta L_\text{min}^{k,c}(\teta^*)$ now defines the $-2\ln$ of the profile likelihood ratio with one parameter ($\teta_k$), and thus by Wilks' theorem the distribution of $\Delta L_\text{min}^{k,c}(\teta^*)$ approaches the chi-square distribution with one degree of freedom in the limit where the data sample is very large. If we now recover the continuum limit by taking $K \rightarrow \infty$, we see that $\Delta L_\text{min}^{k,c}(\teta^*)$ approaches $\Delta L_\text{min}^{c}(v^*,\teta^*)$. Thus the construction of the confidence band is equivalent to finding the collection of confidence intervals in $\teta^*$ for each $v^*$ at a given CL for which $\Delta L^c_\text{min} < \Delta L^*$. Assuming that $\Delta L_\text{min}^c$ is chi-square distributed, the choices $\Delta L^* = 1.0$ and $\Delta L^* = 2.7$ correspond to the confidence intervals of $\teta$ at $68\%$ and $90\%$ CL, respectively, for each $\vmin$ value. In~\cite{Gelmini:2015voa} it was shown that the constrained best fit halo function $\teta_{BF}^c$ defining $L^c_\text{min}(v^*,\teta^*)$ is a piecewise constant function with at most $N_O+1$ steps, with the additional step potentially appearing at $(v^*,\teta^*)$. An in-depth discussion of the interpretation of the confidence band constructed from the profile likelihood ratio is provided in \cite{Gelmini:2015voa}.

\section{Extension of EHI analysis to a global maximum likelihood \label{sec:Extension}}

In this paper we extend the analysis presented in \cite{Gelmini:2015voa} to make statistically meaningful statements about the data of multiple experiments in a halo-independent manner. Specifically, we ($i$) extend the formalism of constructing a pointwise confidence band from a profile likelihood in halo-independent parameter space to a global likelihood function (this section), and ($ii$) propose a method for creating plausibility regions, constructed from a new family of test statistics which can assess the compatibility of multiple data sets under the assumption that the halo function $\teta(\vmin)$ passes through each $(v^*,\teta^*)$ point (see \Sec{sec:constrained}). To accomplish these tasks one must first understand how to find the best fit halo function and constrained best fit halo function from a global likelihood.

In this section we extend the procedure of \cite{Gelmini:2015voa} to the global likelihood function, defined by the product of some number $N_\text{exp}$ of individual likelihood functions, $\alpha =1,2,...N_\text{exp}$,
\begin{equation}
\mathcal{L}_\text{G} = \prod_{\alpha=1}^{N_\text{exp}} \mathcal{L}_{\alpha} \, .
\end{equation}
The procedure of \cite{Gelmini:2015voa} relies on the fact that an extended likelihood function is maximized by a non-increasing piecewise constant $\teta_{BF}(\vmin)$ function with a finite number of points of discontinuity. As discussed below, the methods and reasoning of \cite{Gelmini:2015voa} extend to a global likelihood, if it includes at least one extended likelihood. Thus, the global likelihood function we will work with for the remainder of the paper is
\begin{equation}\label{globallikelihood}
\mathcal{L}_\text{G} = \mathcal{L}_\text{EHI}\prod_{\alpha=1}^{(N_\text{exp}-1)}\mathcal{L}_{\alpha} \, ,
\end{equation}
where $\mathcal{L}_\text{EHI}$ is an extended likelihood (EHI stands for ``extended halo-independent'' \cite{Gelmini:2015voa}) as in \Eq{eq:extendedlikelihood} and, for each $\alpha$, $\mathcal{L}_\alpha$ represents Poisson likelihoods,
\begin{equation}\label{pois}
\mathcal{L}_{\alpha}[\teta] = \prod_{j=1}^{N_\text{bin}^{(\alpha)}} \frac{(\nu_j^{(\alpha)}[\teta] + b_j^{(\alpha)})^{n_j^{(\alpha)}} e^{-(\nu_{j}^{(\alpha)}[\teta]+b_{j}^{(\alpha)})}}{n_j^{(\alpha)}!} \, ,
\end{equation}
or Gaussian likelihoods 
\begin{equation}\label{gaus}
\mathcal{L}_{\alpha}[\teta] = \prod_{j=1}^{N_\text{bin}^{(\alpha)}} \frac{1}{\sigma_j^{(\alpha)} \sqrt{2 \pi}}\exp\left[{-\left(\frac{\nu_j^{(\alpha)}[\teta] + b_j^{(\alpha)} - n_j^{(\alpha)}}{\sqrt{2} \sigma_j^{(\alpha)}}\right)^2}\right] \, .
\end{equation}
Here $\nu_{j}^{(\alpha)}[\teta]$, $b_{j}^{(\alpha)}$, and $n_\text{j}^{(\alpha)}$ are respectively the expected number of dark matter events, the expected number of background events, and the number of observed events in bin $j$ of experiment $\alpha$. $N_{\text{bin}}^{(\alpha)}$ is the number of bins used in the Poisson or Gaussian likelihood of experiment $\alpha$, and $\sigma_j^{(\alpha)}$ is the standard deviation associated with the measurement of $n_j^{(\alpha)}$ in an experiment $\alpha$ employing a Gaussian likelihood.

We now prove that global likelihoods of the form \Eq{globallikelihood} are maximized by non-increasing piecewise constant $\tilde{\eta}$ functions with at most $\mathcal{N}$ steps,
\begin{equation}\label{numsteps}
\mathcal{N} \equiv N_\text{EHI} +\sum_\alpha N^{(\alpha)}_{\rm bin} \, ,
\end{equation}
where $N_\text{EHI}= N_O$ in \Eq{eq:extendedlikelihood}, \ie the number of observed events in the extended likelihood. 

The KKT conditions in Eq.~(\ref{eq:qvmin1}--\ref{eq:qvmin4}) apply equally to any likelihood function $\mathcal{L}$. The KKT condition in \Eq{eq:qvmin4} implies that $\teta_{BF}$ is constant in an open interval where $q(\vmin) \neq 0$. Thus if the $q(\vmin)$ function given by \Eq{eq:qvmin1} has only a  finite number of isolated zeros within a range, the best fit $\teta$ in this range should be a piecewise constant function with steps located at the zeros of $q(\vmin)$. Therefore, the problem of determining the potential number of steps of $\teta_{BF}$ is equivalent to counting the maximum possible number of isolated zeros of the $q(\vmin)$ function.

For the global likelihood in \Eq{globallikelihood}, $q(\vmin)$ is given by
\begin{equation}\label{littleq}
q(\vmin) = 2 \xi^{\text{EHI}}(\vmin) - 2 \sum_{a = 1}^{N_\text{EHI}} \frac{H_a^\text{EHI}(\vmin)}{\gamma_a^\text{EHI}[\tilde{\eta}]} + \sum_{\alpha=1}Q^{(\alpha)}[\tilde{\eta}; \vmin] \, ,
\end{equation}
where $Q^{(\alpha)}[\tilde{\eta}; \vmin]$ is defined by either
\begin{equation}\label{Qpoisson}
Q^{(\alpha)}[\tilde{\eta},\vmin] \equiv \int_{v_\delta}^{\vmin}\, \ud v \, \frac{\delta (- 2 \ln \mathcal{L}_{\alpha})}{\delta \teta(v)} = 2 \sum \limits_{j=1}^{N_\text{bin}^{(\alpha)}} \left[\frac{\nu_j^{(\alpha)}[\tilde{\eta}] + b_j^{(\alpha)} - n_j^{(\alpha)}}{\nu_j^{(\alpha)}[\tilde{\eta}] +b_j^{(\alpha)}}\right] \xi^{(\alpha)}_j(\vmin) \, 
\end{equation}
for Poisson likelihoods of the form in \Eq{pois}, and 
\begin{equation}\label{Qgauss}
Q^{(\alpha)}[\tilde{\eta},\vmin] = 2 \sum \limits_{j=1}^{N_\text{bin}^{(\alpha)}}\left[\frac{\nu_j^{(\alpha)}[\tilde{\eta}] + b_j^{(\alpha)} - n_j^{(\alpha)}}{\sigma_j^2}\right] \xi^{(\alpha)}_j(\vmin) \, ,
\end{equation}
for Gaussian likelihoods in \Eq{gaus}. Changing the function $\teta(\vmin)$ only alters the sign and magnitude of the prefactor of $\xi^{(\alpha)}_j(\vmin)$ in each term of $Q^{(\alpha)}[\teta, \vmin]$. The $\vmin$ dependence of $Q^{(\alpha)}[\teta, \vmin]$ exclusively appears in the functions $\xi^{(\alpha)}_j(\vmin)$, which is defined as in \Eq{eq:xi}, replacing the integration range $[\Ed_\text{min},\Ed_\text{max}]$ with the energy range of the bin, and $\mathcal{H}$ by $\mathcal{H}^{(\alpha)}$. The function $\xi_j^{(\alpha)}(\vmin)$ has the same generic behavior as $\xi(\vmin)$ described at the end of Sec.~\ref{ehianalysis}.

In \App{app:zeros} we prove that above a certain value of $\vmin$, given by the minimum $v_\text{low}^{\mu}$ (see \App{app:zerosa1} for definition), the zeros of $q(\vmin)$ in \Eq{littleq} are isolated, and the maximum number of isolated zeros is given by \Eq{numsteps}. However, in practice the number of steps is smaller than $\mathcal{N}$ and can be determined by studying the functional form of the functions $\xi^\text{EHI}(\vmin)$, $H_a^\text{EHI}(\vmin)$, and $\xi^{(\alpha)}_j(\vmin)$ (which are independent of $\teta$). In Appendix B we prove the uniqueness of the best fit  halo function, $\teta_{BF}$.

\begin{figure*}
\center
\includegraphics[width=.5\textwidth]{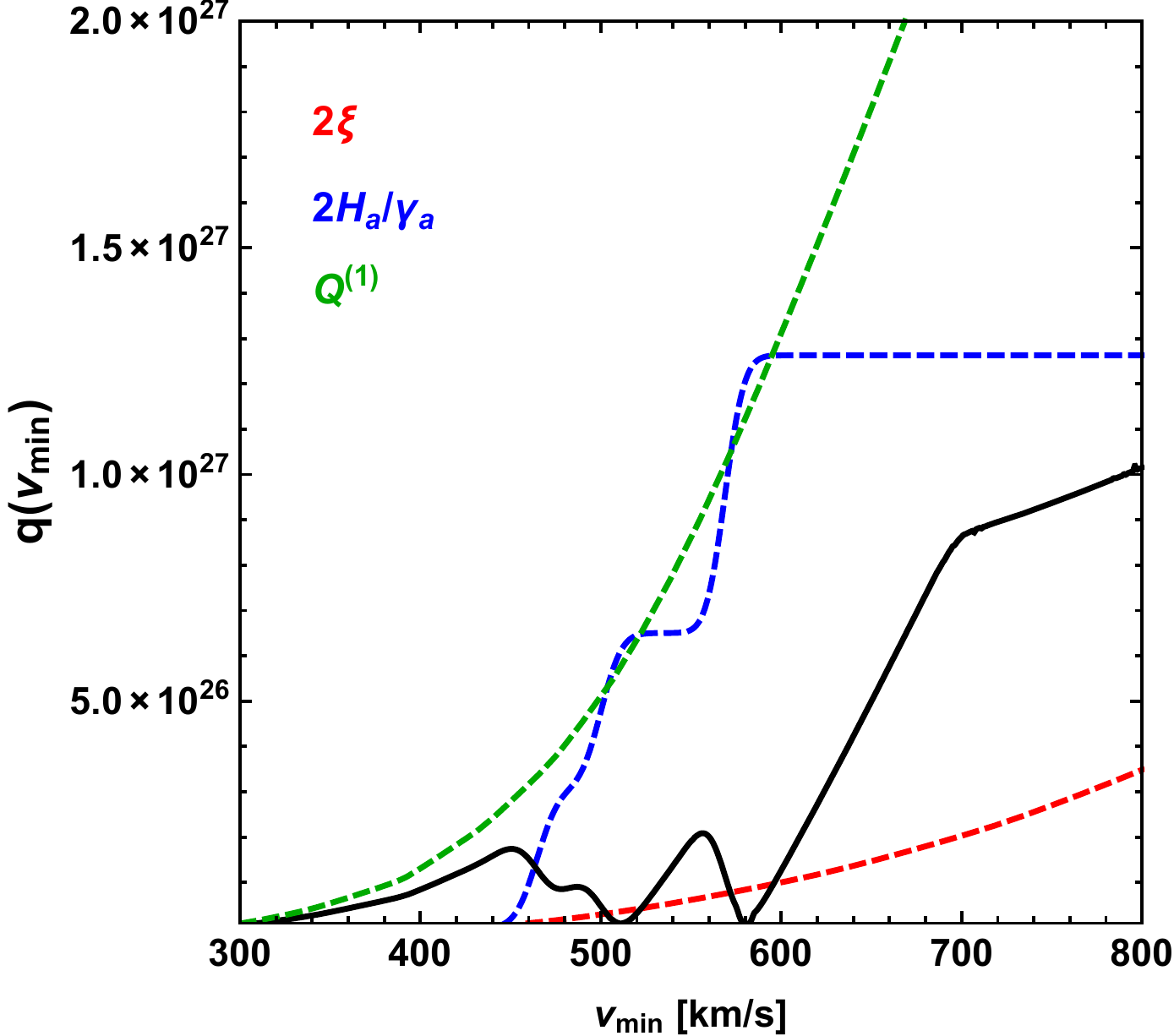}
\caption{\label{fig:qplot}The function $q(\vmin)$ (black), and  absolute value of each of its three comprising terms in \Eq{littleq} (dashed lines) for the combined analysis of CDMS-II-Si and SuperCDMS, assuming a $9 \GeV$ DM particle scattering elastically with a SI contact interaction and $f_n/f_p=1$. The $\vmin$ values where $q(\vmin) = 0$ correspond to the locations of the steps in the global $\teta_{BF}$ halo function. }
\end{figure*}

\begin{figure*}
\center
\includegraphics[width=.49\textwidth]{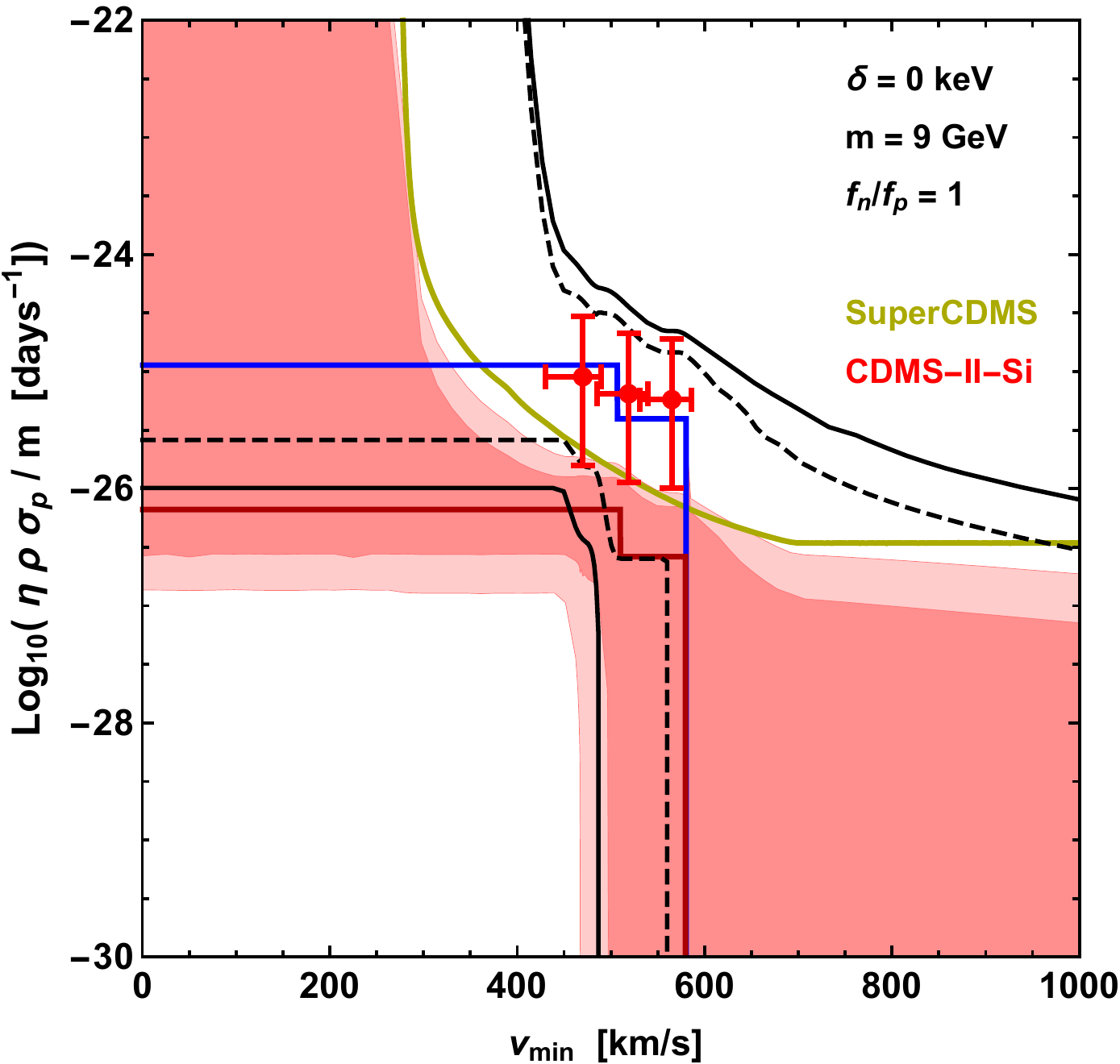}
\includegraphics[width=.49\textwidth]{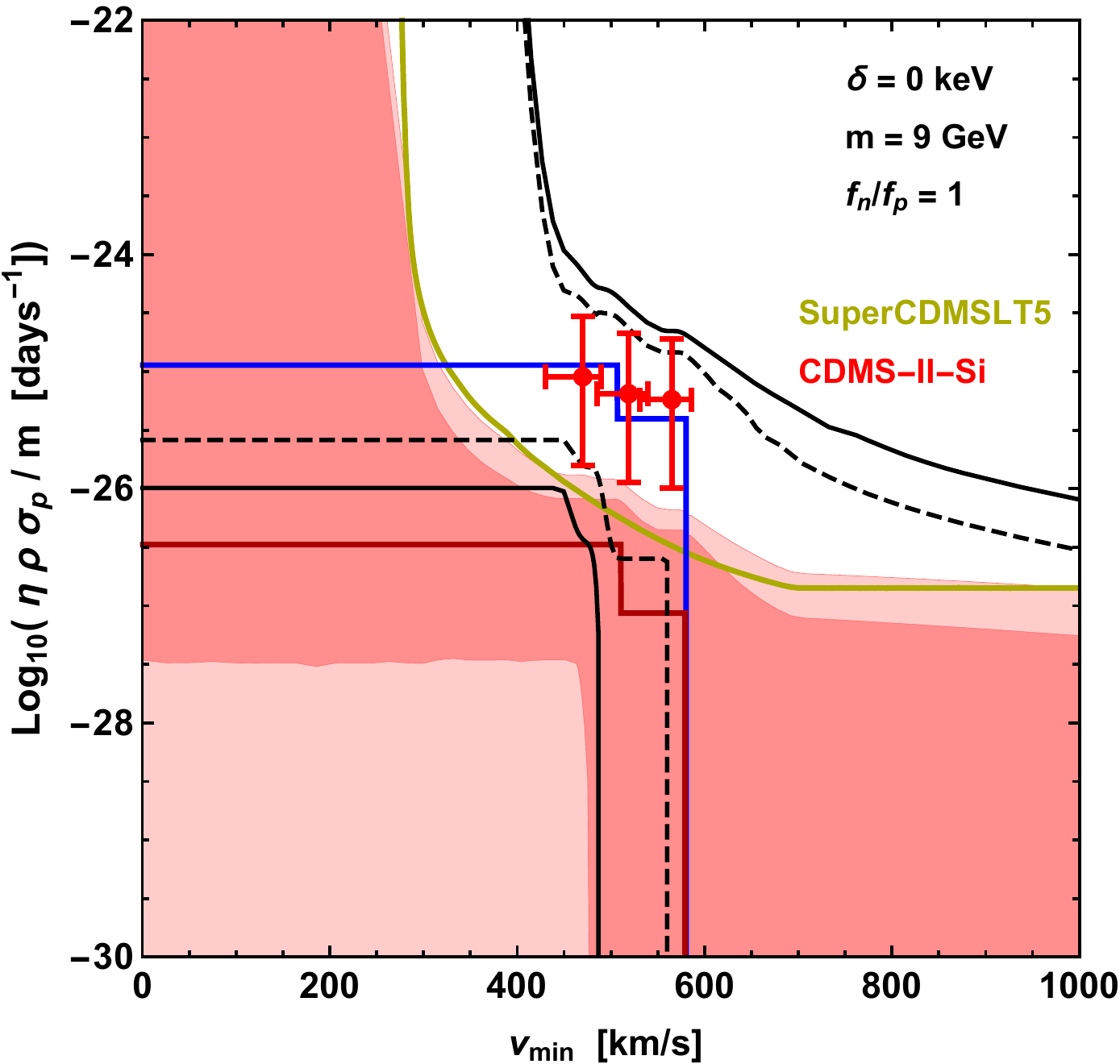}
\caption{\label{fig:supercdms_full}
 68\% (dark red region) and 90\% (light red region) global confidence bands and $\teta_\text{BF}$ (dark red) arising from the combined CDMS-II-Si and SuperCDMS (left) and SuperCDMSLT5 (right) halo-independent analysis. Results are compared with 90\% CL bounds from SuperCDMS (dark yellow line), 68\% (black dashed lines) and 90\% (solid black lines) confidence bands and $\teta_\text{BF}$ for CMDS-II-Si only analysis (blue line) \cite{Gelmini:2015voa}. The red crosses represent the $68\%$ CL intervals of the averaged $\teta$ arising from binning the CDMS-II-Si events into $2$ keVnr bins between $7$ and $13$ keV (see \eg \cite{DelNobile:2013gba,DelNobile:2013cta,DelNobile:2014sja}, in which we take the horizontal bars to be the $\vmin$ range where $90\%$ of the area under $\mathcal{R}_{[\Ed_1,\Ed_2]}(\vmin)$ is contained). The results shown assume a $9 \GeV$ DM particle scattering elastically through a SI isospin-conserving contact interaction ($f_n/f_p = 1$).}
\end{figure*}

An explicit example of the $q(\vmin)$ function and its components is shown in \Fig{fig:qplot} for the case of CDMS-II-Si combined with SuperCDMS data. For SuperCDMS we have taken a one-bin Poisson likelihood, summing over all detectors in Table 1 of \cite{Agnese:2014aze}, the contribution from which to $q(\vmin)$ is shown in green. Also included in \Fig{fig:qplot} are the contributions to $q(\vmin)$ arising from $\xi^\text{EHI}(\vmin)$ (red) and the summation over the $H_a(\vmin)/\gamma_a[\teta]$ (blue). \Fig{fig:qplot} shows that $q(\vmin)$ goes to $0$ at $\vmin \simeq 510$ km/s and $580$ km/s, denoting the locations of the steps of $\teta_{BF}$ (shown later in \Fig{fig:supercdms_full}).

We would like to emphasize that all of the aforementioned arguments have relied on having a global likelihood that contains at least one extended likelihood. This likelihood has the essential feature of contributing an $\teta$-dependent term and an $\teta$-independent term to $q(\vmin)$, with different functional dependences on $\vmin$. 

In order to construct the two sided confidence band, we compute at each value of $\vmin = v^*$ the two sided interval defined by
\begin{equation}
\Delta L_\text{G,min}^c \equiv - 2 \ln \left[ \frac{\hat{\mathcal{L}}_G (v^*,\teta^*)}{\hat{\mathcal{L}}_G} \right] \leq \Delta L^* \, ,
\end{equation}
where $\hat{\mathcal{L}}_G (v^*,\teta^*)$ is the maximum of the global likelihood subject to constraint \Eq{constrain}, and $\hat{\mathcal{L}}_G$ is the maximum of the global likelihood. Using the same arguments of Sec.~\ref{sec:confidencebands} and assuming that $\Delta L_\text{G,min}^c$ is chi-square distributed, the distribution of $\Delta L_\text{G,min}^c$ has one degree of freedom and $\Delta L^* = 1.0$ and $\Delta L^* = 2.7$ for the $68\%$ and $90\%$ CL intervals, respectively. In Sec.~4.2 of \cite{Gelmini:2015voa} it was shown that if $\mathcal{L}$ is maximized by an $\teta_{BF}$ function with a maximum of $N$ steps, then $\mathcal{L}(v^*,\teta^*)$ (\ie $\mathcal{L}$ subject to the constraint that $\teta(\vmin)$ passes through the point $(v^*,\teta^*)$) is maximized by a halo function, which we call the constrained best fit $\teta_{BF}^c$, with a maximum of $(N+1)$ steps, one of which could occur at $\vmin = v^*$. This proof applies to $\mathcal{L}_G$ and $\mathcal{L}_G(v^*,\teta^*)$ as well.

\section{Global Likelihood Analysis of CDMS-II-Si and SuperCDMS data \label{globalband1}}
Here we apply the formalism described in Sec.~\ref{sec:Extension} using the global likelihood function in \Eq{globallikelihood} with an extended likelihood~\cite{Barlow:1990vc} for the three events observed by CDMS-II-Si~\cite{Agnese:2013rvf}, and a $1$-bin Poisson likelihood for SuperCDMS~\cite{Agnese:2014aze}. To obtain background estimates for CDMS-II-Si, we take the normalized background distribution functions from~\cite{McCarthy} and rescale them such that $0.41$, $0.13$, and $0.08$ events are expected from surface events, neutrons, and $^{208}$Pb respectively (see \cite{Agnese:2013rvf}). Since the resolution function for silicon in CDMS-II has not been measured, we take the energy resolution function for germanium from Eq.1 of~\cite{Ahmed:2009rh}. 

In addition to implementing the full SuperCDMS data in Table 1 of \cite{Agnese:2014aze} ($11$ events observed, $6.56$ expected background events, 577 kg-days of exposure), we also use a subset of the SuperCDMS data which neglects the observed events (and the exposure) from tower 5 (4 events observed, 5.33 expected background events, 412 kg-days of exposure). The SuperCDMS collaboration acknowledges that tower 5 had a malfunctioning guard electrode which resulted in a poor understanding of the background in this tower. We will use the label ``SuperCDMSLT5'' for this analysis (where LT5 stands for ``Less Tower 5''). 

The data analysis used throughout this paper is included in the CoddsDM software~\cite{Codds}, an open-source Python program for the analysis of dark matter direct detection data.

In the left panel of Fig.~\ref{fig:supercdms_full} we show the $68\%$ (dark red) and $90\%$ (light red) CL confidence bands, calculated assuming $\Delta L_\text{min}^{c}(v^*,\teta^*)$ is $\chi^2$ distributed with one degree of freedom, for the combined analysis of CDMS-II-Si and SuperCDMS, assuming a $9 \GeV$ DM particle scattering elastically off nuclei with a SI isospin-conserving contact interaction. Also shown in Fig.~\ref{fig:supercdms_full} is the global $\teta_{BF}$ function (dark red line), the $\teta_{BF}$ function for CDMS-II-Si data alone (blue line), the SuperCDMS $90 \%$ upper limit (dark yellow), and the upper and lower boundaries of the $68\%$ (black dashed) and $90\%$ (black solid) CL confidence bands obtained using CDMS-II-Si data alone (these coincide with those presented in Fig.~3 of \cite{Gelmini:2015voa}). Notice that the confidence bands are unbounded from above for $\vmin \lesssim 275$ km/s and $\vmin \lesssim 400$ km/s, for the global analyses and CDMS-II-Si analyses respectively (the lower boundaries of the confidence bands are, however, well defined as $\teta(\vmin)$ is a non-increasing function). This is because $q(\vmin)=0$ in these intervals (\ie the experiment/experiments are not sensitive to recoils imparted from DM traveling at these speeds), and thus the $\teta_{BF}$ is actually undetermined. Since the purpose of plotting these functions is to compare the compatibility of putative and null signals, we extend $\teta_{BF}$ in our plots to this region, in the most conservative way (\ie constant). The red crosses in Fig.~\ref{fig:supercdms_full} represent the $68\%$ CL intervals (vertical bars) of averaged $\teta$ over corresponding $\vmin$ intervals (indicated by horizontal bars) arising from binning the CDMS-II-Si events into $2$ keVnr bins between $7$ and $13$ keV (see \eg \cite{DelNobile:2013gba,DelNobile:2013cta,DelNobile:2014sja}, except we take the horizontal bars to be defined by the $\vmin$ range where $90\%$ of the area under $\mathcal{R}_{[\Ed_1,\Ed_2]}(\vmin)$ is contained). 
 
To determine the all upper bounds on $\teta^0$ arising throughout this paper from the SuperCDMS data, we follow the procedure first outlined in~\cite{Fox:2010bz,Frandsen:2011gi}. Using the fact that $\teta^0(\vmin)$ is a non-increasing function, this procedure argues the smallest possible function passing through a point $(v_0,\teta_0)$ is the downward step-function $\teta_0 \Theta(v_0-\vmin)$. With this in mind, \Eq{eq:rateeq} can be rewritten such that an upper bound on the observed rate in the energy range $[\Ed_1,\Ed_2]$ can be translated into an upper bound $\teta^\text{lim}(\vmin)$ on $\teta^0$, using
\begin{equation}
\teta^\text{lim}(v_0) = \frac{R^\text{lim}_{[\Ed_1,\Ed_2]}}{\int_{v_\delta}^{v_0}\ud \vmin \mathcal{R}_{[\Ed_1,\Ed_2]}(\vmin)} \, .
\end{equation}
This limit is conservative in that every $\teta^0$ function lying above the bound is excluded by the data, but not all $\teta^0$ functions lying below the bound are allowed by the data. The values of $R^\text{lim}$ used in this paper are determined using the Feldman-Cousins approach~\cite{Feldman:1997qc}. Assuming a Poisson distribution for both SuperCDMS ($n=11$, $b=6.56$) and SuperCDMSLT5 ($n=4$, $b=5.33$) and an energy range $[\Ed_1,\Ed_2]$ corresponding to the quoted experimental range (\ie $\Ed_1 = 1.6$ keV and $\Ed_2= 10.0$ keV), this leads to $90\%$ CL upper limits on the number of DM events $\mu^\text{lim}$ of $11.25$ and $3.33$ events respectively. The value of $R^\text{lim}$ can then be obtained by dividing $\mu^\text{lim}$ by the exposure of the relevant experiment.     

The global $\teta_{BF}$ function is shifted to lower values of $\teta$ by over an order of magnitude relative to the $\teta_{BF}$ found using CDMS-II-Si data alone, and is outside the $68\%$ and $90 \%$ CL confidence bands of CDMS-II-Si alone. Similarly, the $\teta_{BF}$ for CDMS-II-Si alone (in blue) is incompatible with the $68\%$ and $90\%$ Cl global confidence bands. Furthermore, in the range $360 \text{ km/s } \lesssim \vmin \lesssim 480$ km/s the $68\%$ CL global confidence band has no overlap with the $68 \%$ CL confidence band of CDMS-II-Si.

The right panel of Fig.~\ref{fig:supercdms_full} is the same as the left panel but using SuperCDMSLT5 instead of SuperCDMS. The global $\teta_{BF}$ function has shifted to slightly lower values of $\teta$ (relative to the SuperCDMS analysis), as have both confidence bands, but the general conclusions are the same -- namely, there appears to be a strong level of incompatibility between the results arising from the global likelihood and those found using only CDMS-II-Si data. We also note that the increased conflict between CDMS-II-Si and SuperCDMSLT5 has resulted in the $90 \%$ CL confidence band extending down to $\teta \simeq 0$ (\ie no DM) at low values of $\vmin$, as opposed to having a well defined non-zero lower boundary for the case of SuperCDMS. 

\begin{figure*}
\center
\includegraphics[width=.5\textwidth]{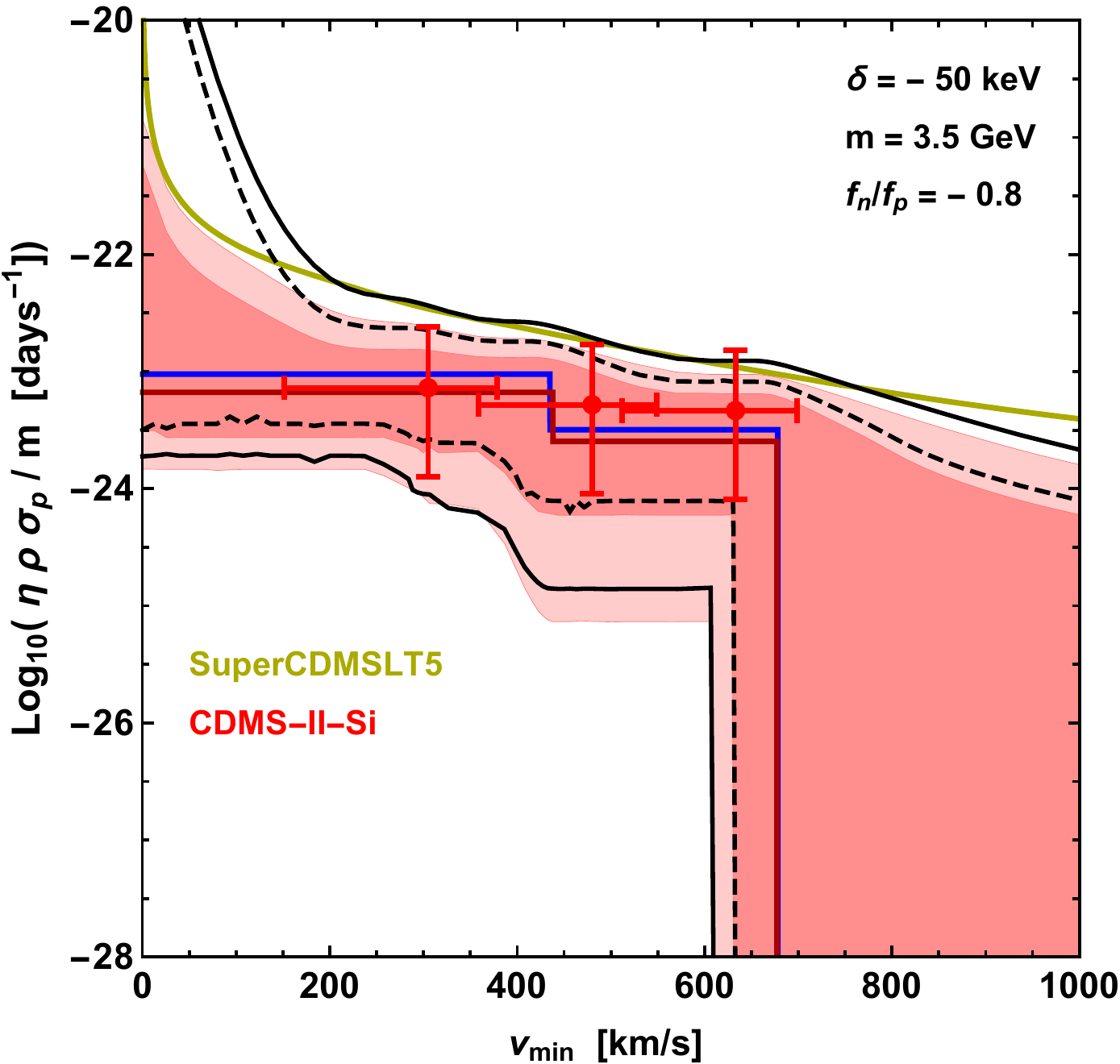}
\caption{\label{fig:gephobic} Same as  the right panel of Fig.~\ref{fig:supercdms_full} but for a $3.5 \GeV$  DM particle with exothermic scattering ($\delta = -50$ keV) and  a  Ge-phobic SI interaction ($f_n/f_p = -0.8$)~\cite{Gelmini:2014psa} }
\end{figure*}

We present one final illustration of this method in Fig.~\ref{fig:gephobic} for a $3.5 \GeV$  DM particle with exothermic scattering ($\delta = -50$ keV) and  a  Ge-phobic SI interaction ($f_n/f_p = -0.8$)~\cite{Gelmini:2014psa}. This example has been chosen to illustrate how the global $\teta_{BF}$ and confidence bands behave in the case of non-conflicting data sets. As expected, the results from the global likelihood analysis of CDMS-II-Si and SuperCDMSLT5 are nearly identical to the results obtained from CDMS-II-Si alone, with the only significant change occurring at low values of $\vmin$, where the upper bound of SuperCDMSLT5 is in conflict with the confidence bands of CDMS-II-Si alone.

\section{Constrained Goodness-of-Fit Analysis \label{sec:constrained}}

The global likelihood analysis presented in the previous section always produces a best fit halo function and confidence band, even when considering conflicting data sets.  A particular goodness-of-fit test has been proposed in \cite{Maltoni:2002xd,Maltoni:2003cu} to assess the compatibility of different data sets in the framework of a given theoretical model. This so called ``parameter goodness-of-fit'' (PG) test was used in~\cite{Feldstein:2014ufa} to gauge the compatibility of CDMS-II-Si, SuperCDMS, and LUX data, in a halo-independent way. It is defined as  
\begin{equation}
q_{PG} \equiv -2 \left(\ln \hat{\mathcal{L}}_\text{G} - \sum_{\alpha} \ln \hat{\mathcal{L}}_{\alpha} \right) \, ,
\end{equation}
where $\hat{\mathcal{L}}_\text{G}$ is the maximum of the global likelihood and $\hat{\mathcal{L}}_{\alpha}$ is the maximum of the likelihood of experiment $\alpha$.
If the $\teta_{BF}$ of all individual experiments would coincide, then $q_{PG}=0$. On the other hand a strong disagreement between the $\teta_{BF}$ of individual experiments would lead to a large value of $q_{PG}$. Thus $q_{PG}$ quantifies the degree of compatibility of all data sets under the assumption of a particular DM particle model. To provide a quantitative statement about the compatibility, the $p$-value of the observed data was obtained from a MC simulation, assuming the global $\teta_{BF}$ is the true  halo model~\cite{Feldstein:2014ufa}. This procedure assigns a single number, a single $p$-value, to the whole halo-independent parameter space, and we would like to identify regions of this  space where  $\teta(\vmin)$ functions may lead to better or worse compatibility among data sets. With this purpose in mind, we define a family of test statistics similar to $q_{PG}$, one for each point in parameter space, using the profile likelihood, defined as the likelihood maximized subject to the constraint in \Eq{constrain}, \ie $\teta(v^*)=\teta^*$ (it is the continuum limit of the numerator inside the square bracket in \Eq{consproflike}). We will then define a $p$-value for every point in the halo independent parameter space. We define the ``constrained parameter goodness-of-fit'' test statistic as
\begin{equation}
q_{PG}^c(v^*,\eta^*) \equiv -2 \left(\ln \hat{\mathcal{L}}_\text{G}^c(v^*,\teta^*) - \sum_{\alpha} \ln \hat{\mathcal{L}}_{\alpha}^c(v^*,\teta^*) \right) \, ,
\end{equation}
where $\hat{\mathcal{L}}_{G}^c(v^*,\teta^*)$ is the global profile likelihood and $\hat{\mathcal{L}}_{\alpha}^c(v^*,\teta^*)$ is the profile likelihood of experiment $\alpha$. $q_{PG}^c$ tests the compatibility of the different data sets under the assumption that $\teta(\vmin)$ passes through $(v^*,\teta^*)$. To infer the probability distribution for $q_{PG}^c(v^*,\teta^*)$ we use a Monte Carlo simulation, assuming the true halo model is given by the global best fit halo function that maximizes $\mathcal{L}_\text{G}$ under the constraint $\teta(v^*)=\teta^*$. We call  ``constrained best fit halo function" $\teta^c_{BF}$ the function that maximizes a likelihood subjected to this constraint. There is a different  $\teta^c_{BF}(\vmin)$  function for each $(v^*, \teta^*)$ point (which certainly fulfills the condition $\teta^c_{BF}(v^*) =\teta^*$), one for the global likelihood and one for each single experiment extended likelihood.  The $p$-value for a given $(v^*,\teta^*)$ is then obtained by comparing the observed value of $q_{PG}^c$ to the distribution constructed from $\Ord(10^3)$ simulated data sets (for each choice of $(v^*,\teta^*)$).

\begin{figure*}
\includegraphics[width=.5\textwidth]{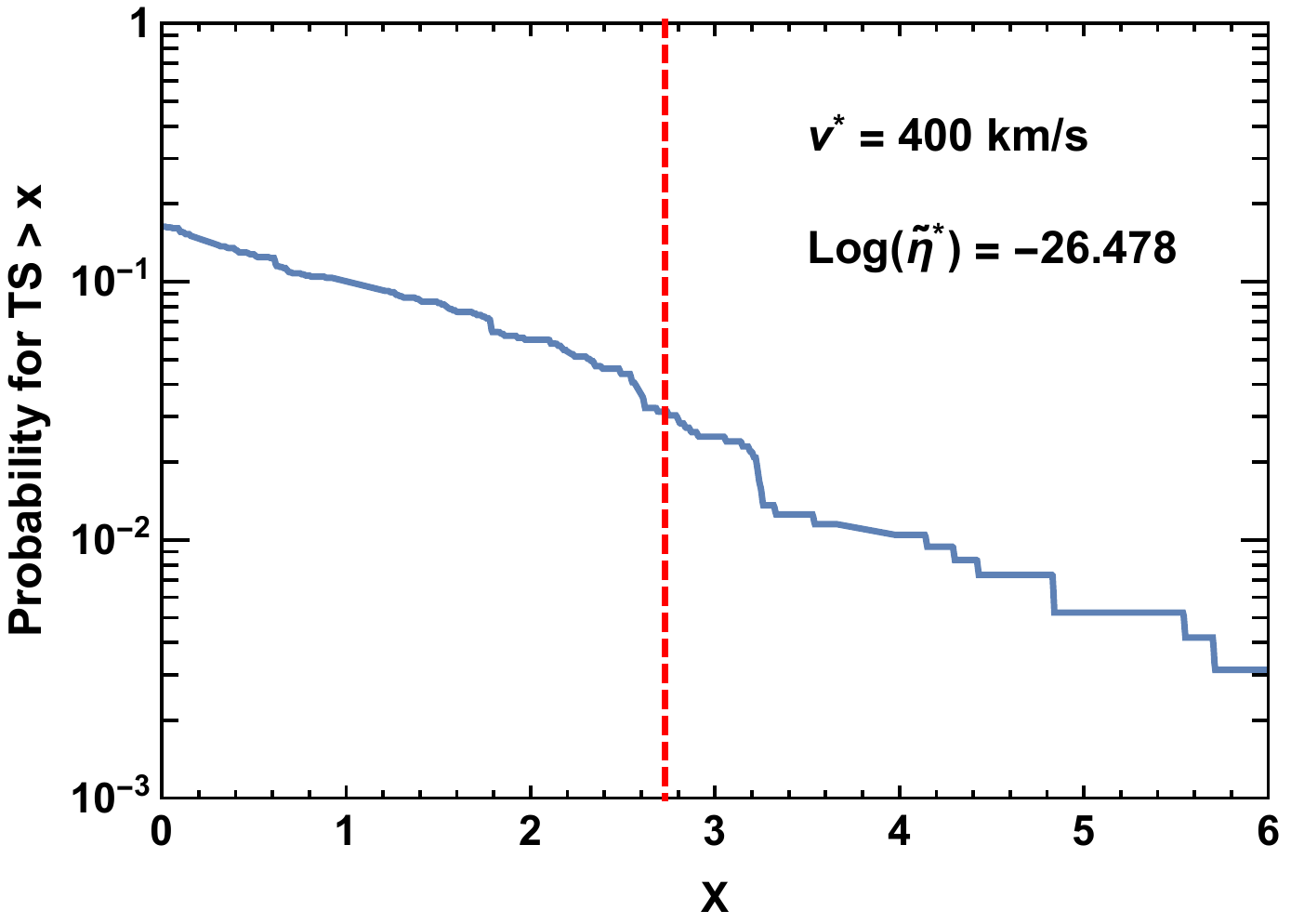}
\includegraphics[width=.5\textwidth]{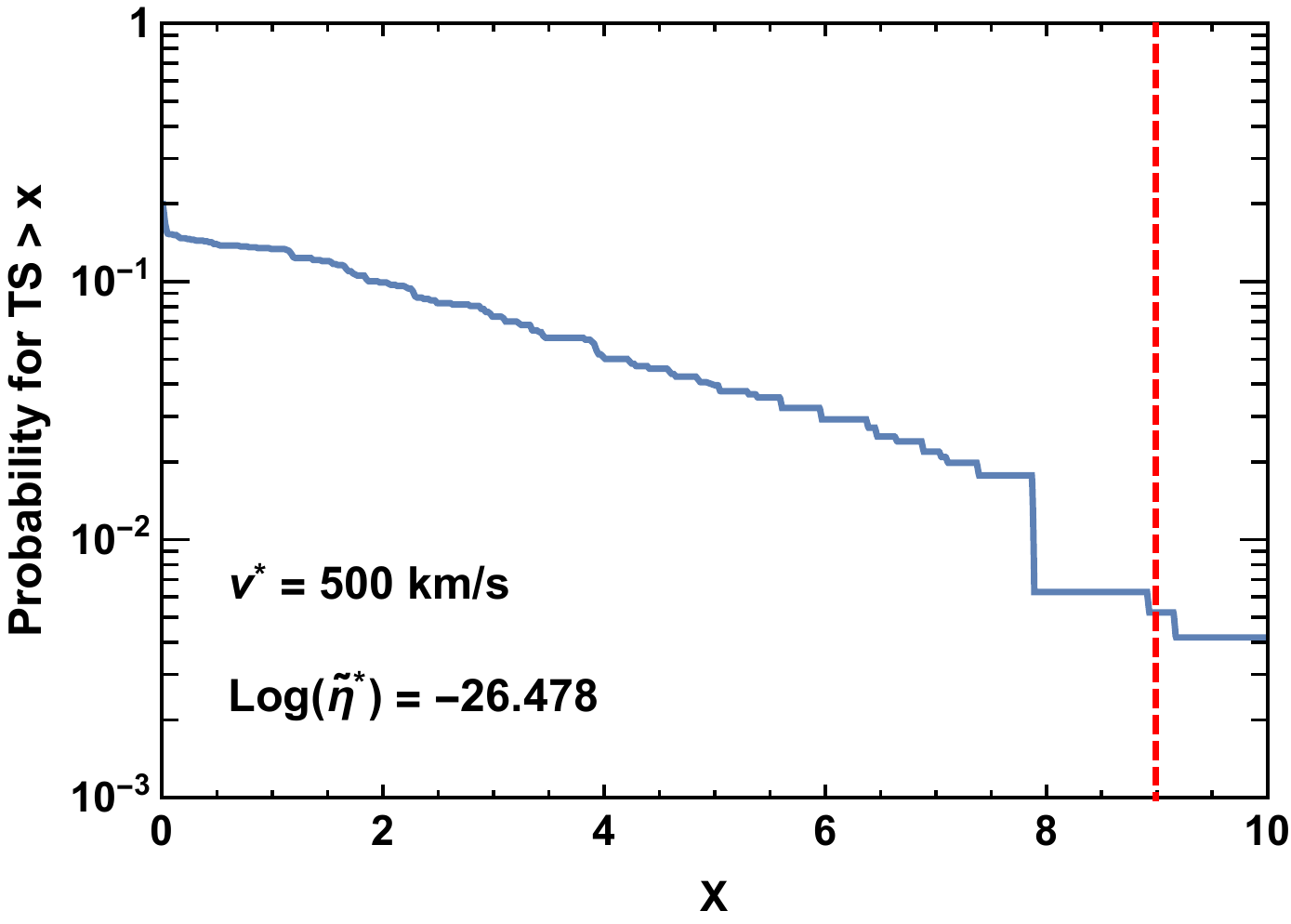}
\caption{{\label{fig:MC_check}} Monte Carlo generated distribution for $q_{PG}^{c}$ for $v^* = 400$ km/s (left) and $500$ km/s (right) and $\log(\teta^*) = -26.478$. Observed value of $q_{PG}^c$ shown with red dashed line.}
\end{figure*}

We have only developed a method for maximizing the Poisson and Gaussian likelihoods subject to the constraint $\teta(v^*) = \teta^*$ for a single bin Poisson/Gaussian likelihood. In this case, the likelihood is maximized by an expected number of dark matter events $\hat{\nu}_1^{(\alpha)}$, where either $\hat{\nu}_1^{(\alpha)} = n_1^{(\alpha)}-b_1^{(\alpha)}$ if $n_1^{(\alpha)}\geq b_1^{(\alpha)}$, or $\hat{\nu}_1^{(\alpha)} = 0$ if $n_1^{(\alpha)}\leq b_1^{(\alpha)}$. In order to maximize the constrained likelihood, one needs to consider whether $v^*$ lies above or below the experimental threshold. If $v^*$ is below threshold, a halo function passing through $(v^*,\teta^*)$ produces a minimum number of $0$ observed events (with $\teta = \teta^* \Theta (v^*-\vmin)$), and a maximum number $\nu_\text{max}$ of events given by the flat halo function $\teta(\vmin)=\teta^*$. If $v^*$ is above threshold, a halo function passing through $(v^*,\teta^*)$ produces a minimum number $\nu_\text{min}$ of observed events when $\teta = \teta^* \Theta (v^*-\vmin)$, and there is no limit on the maximum number of observed events because $\teta$ can be unbounded from above for $\vmin < v^*$. If $\hat{\nu}_j^{(\alpha)}$ lies between the minimum and maximum number of predicted events for $\teta(\vmin)$ passing through $(v^*,\teta^*)$ in each case, then the maximum of the constrained likelihood is the maximum of the likelihood. Otherwise, the maximum of the constrained likelihood is calculated using $\nu_\text{max}$ or $\nu_\text{min}$, depending on the respective case above.

The probability distributions of $q_{PG}^c$ are shown in Fig.~\ref{fig:MC_check} for the combination of CDMS-II-Si and SuperCDMSLT5, for a SI contact interaction with $(m, \delta, f_n/f_p) = (9 \, \GeV, 0 \, \text{keV}, 1)$, for $v^* = 400$ km/s (left) and $500$ km/s (right) with $\teta^*$ chosen on the global $\teta_{BF}$ curve. The observed value of $q_{PG}^c$ in Fig.~\ref{fig:MC_check} are indicated by the dashed red line. The $p$-values roughly correspond to $2.8 \%$ for $v^* = 400$ km/s, and $0.5 \%$ for $v^* = 500$ km/s. While the probability distributions shown in Fig.~\ref{fig:MC_check} do not appear to approach $1$ in the limit $x \rightarrow 0$, there are in fact a large number of simulations which yield extremely small values of $q_{PG}^c$ that are not depicted (the probabilities do in fact equal $1$ at $x=0$). This happens because the global best fit halo function predicts less than one observed event in CDMS-II-Si, which leads to many simulations in which $0$ events are observed by CDMS-II-Si. In turn, this implies the global constrained best fit halo function and the constrained best fit halo function for CDMS-II-Si are the same, as they can only have a single step at the location of $(v^*,\teta^*)$. For SuperCDMSLT5, the expected background is larger than the number of observed events, and thus the profile likelihood of SuperCDMSLT5 is relatively insensitive to halo functions that predict small numbers of DM events. Consequently, it is not uncommon to find $\ln\hat{\mathcal{L}}_\text{G}^c(v^*,\teta^*) \simeq \sum_{\alpha} \ln\hat{\mathcal{L}}_{\alpha}^c(v^*,\teta^*) $.

\begin{figure*}
\center
\includegraphics[width=.49\textwidth]{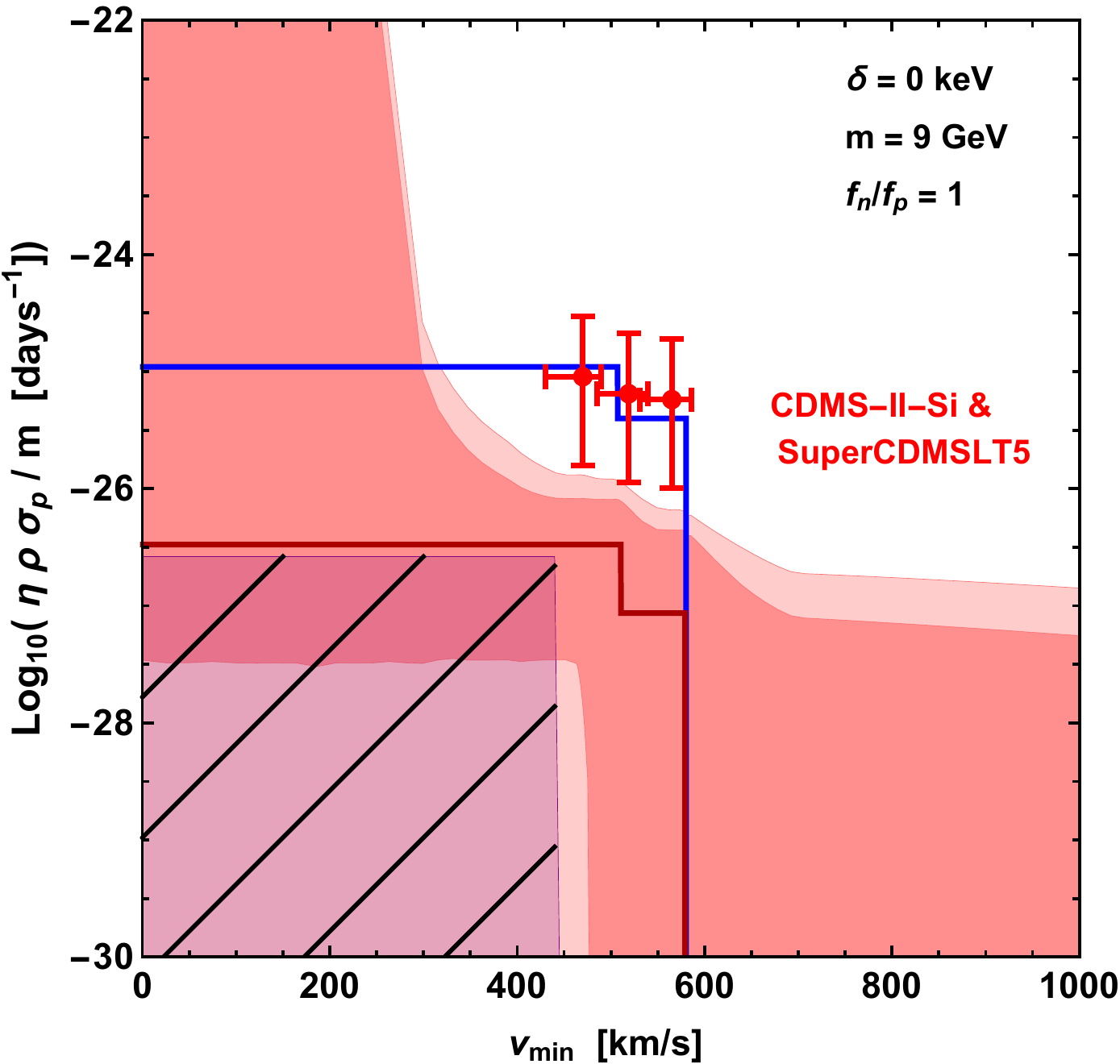}
\includegraphics[width=.49\textwidth]{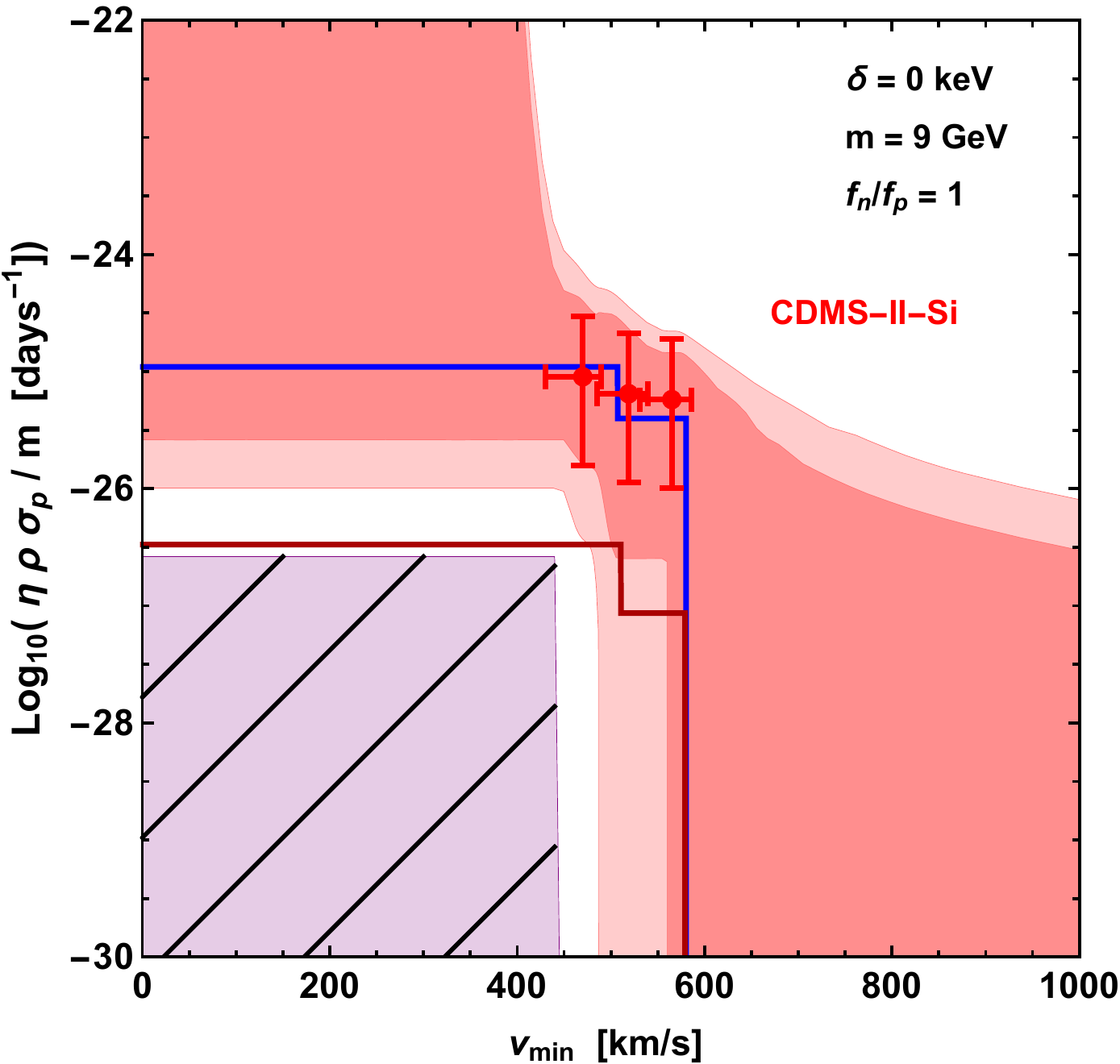}
\caption{\label{fig:MCband} Plausibility region (light purple) generated from the constrained parameter goodness-of-fit test statistic for CDMS-II-Si and SuperCDMSLT5 ($p$-value larger than $10\%$), compared with the confidence bands (red shaded) generated for CDMS-II-Si data alone (left)~\cite{Gelmini:2015voa} and the global confidence bands (red shaded) constructed in Sec.~\ref{globalband1} (right). The plausibility regions are crossed over because halo functions entirely contained within these regions are not necessary allowed by our test, i.e. do not necessarily lead to a compatibility of the data sets at the level of $p > $10\%. However, for any halo function not entirely contained within the plausibility region the data sets are incompatible at the chosen level ($p < $10\%). Also shown are $\teta_{BF}$ for CDMS-II-Si alone (blue), the $\teta_{BF}$ resulting from the global likelihood analysis (dark red), and the $\vmin$-averaged CDMS-II-Si data (crosses) as described in Sec.~\ref{globalband1}.}
\end{figure*}

\begin{figure*}
\center
\includegraphics[width=.49\textwidth]{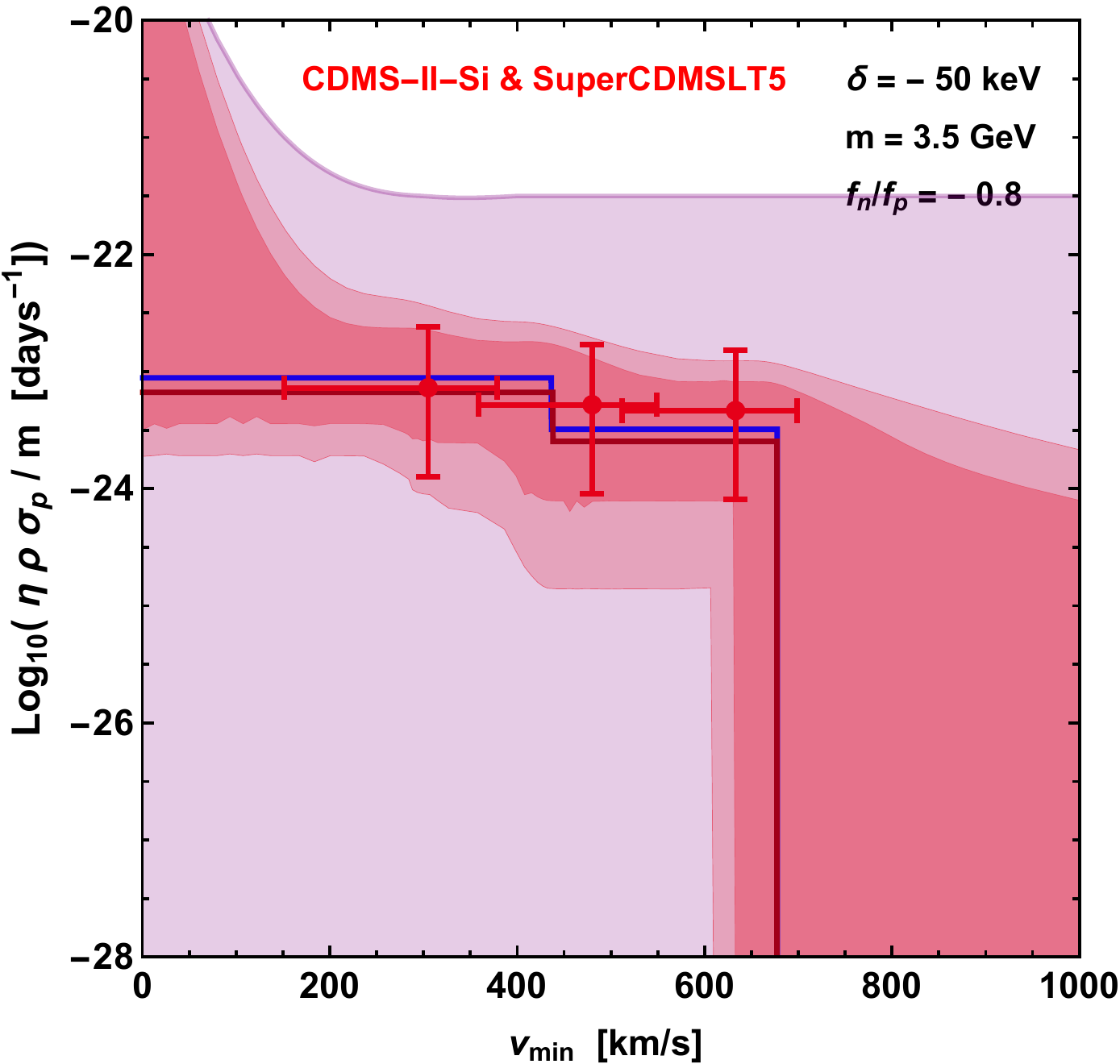}
\caption{\label{fig:MCband-gephobic}  Same as the right panel of Fig.~\ref{fig:MCband} but for a $3.5 \GeV$  DM particle with exothermic scattering ($\delta = -50$ keV) and  a  Ge-phobic SI interaction ($f_n/f_p = -0.8$)~\cite{Gelmini:2014psa} (as in Fig.~\ref{fig:gephobic}). Halo functions $\teta{\vmin}$ entirely contained within the plausibility region (light purple) lead to a compatibility of the data sets at the chosen level ($p > $10\%). For those not entirely contained within the plausibility region the data sets are incompatible at the chosen level ($p < $10\%). }
\end{figure*}

Fig.~\ref{fig:MC_check} already demonstrates a high level of incompatibility between the CDMS-II-Si and SuperCDMSLT5 data sets for the assumed WIMP candidate, because the global $\teta_{BF}(\vmin)$ cannot produce a large $p$-value, say larger than 10\%. We can construct intervals at each $\vmin=v^*$ in which the probability of obtaining a $q_{PG}^c$ value larger than the one observed is $\geq 10 \%$. By joining these intervals we build regions in $(\vmin, \teta)$ which are referred to as ``plausibility'' regions. 

Let us now clarify the meaning of the plausibility regions. A halo function $ \teta(\vmin)$ is a non-increasing continuous function which must be defined for any value of $\vmin$.  Consequently,
any halo function not entirely contained within the plausibility region passes though points with $p<$ 10 \%, and thus for these functions the data sets are incompatible at the chosen level 
($p <$ 10\%). However, halo functions that are entirely contained within a plausibility region are not necessarily allowed by our test, i.e. do not necessarily lead to compatibility of all data at the chosen level. The issue is that the true halo model adopted at each point within a plausibility region, namely the  $\teta^c_{BF}$ of the profile global likelihood at each point, may also pass through points outside the plausibility region and be rejected by our test. If so, the  $p$-value evaluation at the particular point in the plausibility region is inconsistent. This is the case for all points in the plausibility regions (light purple) shown in Fig.~\ref{fig:MCband}. The regions are crossed by thin black lines to indicate that halo functions  entirely contained within them  are not guaranteed to lead to compatibility of the data sets. However, if the true halo model adopted at all points within a plausibility region are entirely contained within it, the $p$-value calculation is reliable and halo functions entirely contained with this region are allowed by our test.  This is the case of the plausibility region in Fig.~\ref{fig:MCband-gephobic} (shown in light purple).

The plausibility region for $p \geq 10 \%$ arising from the constrained parameter goodness-of-fit test for the combination of CDMS-II-Si and SuperCDMSLT5 (light purple region) is compared in Fig.~\ref{fig:MCband} to the confidence bands (red shaded regions) generated from the global likelihood described in Sec.~\ref{globalband1} (right panel) and the confidence bands generated with CDMS-II-Si data alone (left panel). Also shown are the global best fit halo function $\teta_{BF}$ (dark red) and the best fit halo function for CDMS-II-Si alone (blue). The left panel of Fig.~\ref{fig:MCband} shows that there does not exist a halo function in the CDMS-II-Si confidence bands that can describe the compatibility of the observed data sets. Not all halo functions contained within the $90\%$ global confidence band in the right panel of \Fig{fig:MCband} are excluded by the plausibility region, but the $68\%$ region is entirely excluded, as is the global best fit halo function. As explained above the plausibility regions are crossed over because the functions entirely included within them are not allowed by our test (while those passing through points outside them are rejected). By contrast, we can see in Fig.~\ref{fig:MCband-gephobic} how the plausibility region includes the entire global confidence bands  (as well as the bands for CDMS-II-Si alone, which in this case are nearly identical, see Fig.~\ref{fig:gephobic}) in the case of non-conflicting data sets. This is the example of  a $3.5 \GeV$  DM particle with exothermic scattering ($\delta = -50$ keV) and  a  Ge-phobic SI interaction ($f_n/f_p = -0.8$)~\cite{Gelmini:2014psa}. The plausibility region  provides in this case a further indication of compatibility of the CDMS-II-Si and SuperCDMSLT5  data sets for this particular DM particle model, besides the near complete overlap of the global and single experiment confidence bands.

A comment is in order regarding \Fig{fig:MCband-gephobic}. While generating the probability distributions at large values of $\teta$ and $v_\text{min}$ (above the 90$\%$ CL band), we found the predicted number of events in both experiments was too large for our computational methods to work. We resorted to using a nearest-neighbor extrapolation at fixed $\vmin$ to generate the probability distributions in this region. We found that in this region of the $\vmin-\teta$ plane, the probability distribution changes slowly with respect to the observed value of $q^c_{PG}$, and thus we believe we obtained a good estimate of the upper boundary of the plausibility region. This extrapolation was only used above the 90$\%$ CL confidence band boundary.

\section{Conclusions \label{conclusion}}
In this paper we have presented two distinct methods to assess the joint compatibility of data sets for a given DM particle model across halo-independent parameter space, using a global likelihood consisting of at least one extended likelihood and an arbitrary number of Gaussian or Poisson likelihoods. We have illustrated these methods by applying them to CDMS-II-Si and SuperCDMS data, assuming WIMP candidates with SI contact interactions.

The first method is a natural extension of the procedure presented in \cite{Gelmini:2015voa}, in which a best fit halo function and pointwise confidence band are constructed from the profile likelihood ratio. Here we  have proven that the best fit halo function $\teta_{BF}$ for the global likelihood we studied is a piecewise constant function with the number of steps at most equal to the number of unbinned data points plus the number of data bins in all the single likelihoods, and argued why in practice the number of steps is smaller than this maximum number (see Section 3 and Appendix A).   A best fit piecewise constant halo function had already been found  in the literature (see~\cite{Feldstein:2014ufa}) for a global likelihood of the type we use, but as a curiosity without any explanation (or proof of uniqueness). In addition to showing how to find the best fit halo function $\teta_{BF}$ and that this function is unique (see Appendix B), here we have shown for the first time how to construct two-sided confidence bands at any CL for the type of global likelihood we studied. As an illustration of the method we have found the best fit halo function and the $68 \%$ and $90 \%$ CL confidence bands  assuming two different choices for the DM particle model parameters $m$, $\delta$, and $f_n/f_p$. The choice of a $9 \GeV$ DM particle scattering elastically ($\delta=0$) with an isospin-conserving coupling ($f_n/f_p=1$) leads to an apparent incompatibility between the observed CDMS-II-Si events and the SuperCDMS upper limit, in agreement with previous published results (see \eg \cite{DelNobile:2014sja,Gelmini:2015voa}). This incompatibility can be assessed by comparing the overlap or lack thereof of the global confidence bands with those of CDMS-II-Si alone. As shown in Fig.~\ref{fig:supercdms_full}, at the $68\%$ CL, it is not possible to find a halo function passing through both confidence bands.   The situation is very different  for a $3.5 \GeV$  DM particle with exothermic scattering ($\delta = -50$ keV) and  a  Ge-phobic SI interaction ($f_n/f_p = -0.8$)~\cite{Gelmini:2014psa}, for which the data sets are compatible. As shown in Fig.~\ref{fig:gephobic} the global and CDMS-II-Si alone confidence bands practically coincide.

The drawback of this method is that it cannot provide a quantitative measurement of the level of incompatibility of the various data sets that comprise the global likelihood. To address this concern, we have proposed in Section 5 a second method in which we construct a ``plausibility region" arising from the global likelihood, using an extension of the parameter goodness-of-fit test~\cite{Maltoni:2002xd,Maltoni:2003cu,Feldstein:2014ufa}, that we refer to as the ``constrained parameter goodness-of-fit'' test. By evaluating the ratio of the global profile likelihood and the product of the individual profile likelihoods (assuming $\teta(v^*)=\teta^*$), a plausibility region can be constructed by grouping together regions of parameter space for which, at each point $(v^*,\teta^*)$, our observed test statistic has a $p$-value \eg $\geq 10\%$. This $p$-value was determined using a probability distribution constructed with Monte Carlo generated data assuming the true halo function is the constrained best fit $\teta_{BF}^c$ of the profile global likelihood, \ie the halo  function that maximizes the global likelihood subject to the constraint $\teta(v^*)=\teta^*$. For any halo function not entirely contained within this plausibility region the data are incompatible for the assumed DM particle model at the assumed level (\eg $p < 10\%$). For halo functions entirely contained within the plausibility region the data sets are compatible at the chosen level only if the contained best fit at each point within the region are also entirely contained within the region. We have demonstrated this method for a $9 \GeV$ DM particle scattering elastically with an isospin conserving coupling and for the aforementioned Ge-phobic particle candidate.  The results are shown in Figs.~\ref{fig:MCband}  and \ref{fig:MCband-gephobic} respectively. In the first case the confidence bands are largely outside the plausibility region, while in the second case the confidence bands are entirely included in the plausibility region and any halo function entirely contained within the plausibility region lead to a compatibility of the data sets at the chosen level ($p >$ 10\%).
 
Together these two methods provide complementary assessments of the compatibility of the data given a particular dark matter model, across the  $\vmin-\teta$ halo-independent parameter space. We expect these tools to prove useful for future direct dark matter searches both to test compatibility of different data sets as to provide a guidance of which type of halo functions provide a better or worse compatibility of all the data.

\section{Acknowledgments} 
The authors thank Paolo Gondolo for many useful discussions. G.G.~acknowledges partial support from the Department of Energy under Award Number DE-SC0009937. JHH is supported by the CERN-Korea fellowship through the National Research Foundation of Korea. GG thanks the CERN Theoretical Physics Department for hospitality during part of the writing of this paper.

\appendix

\section{The zeros of the $q(\vmin)$ function \label{app:zeros}}

We are first going to argue that the zeros of $q(\vmin)$ are only isolated above a certain $\vmin$ range where all terms in the sum defining $q(\vmin)$ are zero. We will then find the maximum possible number of isolated zeros, although the actual number of zeros can be much smaller than the maximum.

\subsection{The zeros are isolated above a certain $\vmin$ value}
\label{app:zerosa1}

The terms defining $q(\vmin)$ in \Eq{littleq} are either positive semidefinite, \eg $\xi^{\text{EHI}}(\vmin)$ and some of the terms proportional to $ \xi^{(\alpha)}_j(\vmin)$, or negative semidefinite, \eg the terms proportional to $H_a^\text{EHI}(\vmin)$ and some of the terms proportional to $ \xi^{(\alpha')}_{j'}(\vmin)$. To facilitate a smooth discussion of the behavior of these functions, let us introduce the label $\mu$, which will be used to denote either a quantity associated with the EHI experiment or an experiment-bin pair $(\alpha,j)$. This way, quantities like $\xi^{\mu}(\vmin)$ can either represent $\xi^\text{EHI}(\vmin)$ or $\xi^{(\alpha)}_j(\vmin)$.

Each term in the sum in  \Eq{littleq} has a different $\vmin$-dependence. In particular the $\xi^\mu(\vmin)$ functions (note the general behavior of $\xi^{(\alpha)}_j(\vmin)$ is identical to that of $\xi(\vmin)$ discussed in Sec.~\ref{ehianalysis}) are zero below certain values of $\vmin$, which we will refer to as $v_\text{low}^{\mu}$, strictly increase with $\vmin$ (although the second derivative may exhibit sign changes), until at some value of $\vmin$, call it $v_\text{high}^{\mu}$, they plateau and become constant. The $v_\text{low}^{\mu}$ and $v_\text{high}^{\mu}$ of each $\xi^{\mu}(\vmin)$ function, as well as the height of the plateau, depend on theoretical framework and the specifics of the experiments (\eg the scattering kinematics, the differential cross section, the energy resolution functions, etc.). The $H_a^\text{EHI}(\vmin)$, also described in Sec.~\ref{ehianalysis}, are instead upward step-like functions, starting from zero at low $\vmin$, with the steps appearing roughly at the $\vmin$ values corresponding to the detected energy of the events observed in the EHI experiment.  

In addition to having unique $\vmin$-dependencies, each of the terms in \Eq{littleq} has uniquely defined $\teta$-dependent coefficient. Thus the terms are all independent of each other and have very different functional forms.

For values of $\vmin$ below the minimum $v_\text{low}^{\mu}$, i.e. where all the terms in \Eq{littleq} are zero, $q(\vmin)$ is zero, which implies $\teta_{BF}(\vmin)$ is undetermined. This is not detrimental to the arguments we have made as it reflects the fact that experiments under consideration do not probe the halo function at these values of $\vmin$. Notice that in order to have non-negative $q(\vmin)$ values, the $v_\text{low}$ of some of the positive terms must be smaller than the smallest $v_\text{low}$ of all negative terms.

For values of $\vmin$ larger than the minimum $v_\text{low}^{\mu}$, zeros of $q(\vmin)$ can appear where the modulus of the sum of all negative terms in \Eq{littleq} touches from below the sum of all positive terms in \Eq{littleq} (recall that $q(\vmin)$ is a non-negative function). The positive terms consist of different $\xi^\mu(\vmin)$ (most of them multiplied by $\tilde{\eta}$-dependent coefficients). Thus, in general, the sum of all positive terms behaves as a monotonically increasing function starting from zero at the lowest $v_\text{low}^{\mu}$ (lowest of all positive terms) and plateauing to a constant value at the largest $v_\text{high}^{\mu}$ (again considering only positive terms). The negative terms in \Eq{littleq} include the step-like $H_a(\vmin)$ (multiplied by $\tilde{\eta}$ dependent coefficients), which each add a ``step-like" feature to the modulus of the sum of negative terms, and some of the $\xi^{(\alpha)}_j(\vmin)$ dependent terms (multiplied by $\tilde{\eta}$ dependent negative coefficients). Depending on the nature of these negative $\xi^{(\alpha)}_j(\vmin)$ terms, they could add ``shoulder-like'' features, arising from changes in the sign of the second derivative, to the modulus of the sum of negative terms. The modulus of the sum of negative terms also plateaus above the largest $v_\text{high}^\mu$ (largest of all negative terms). The plateau of the sum of positive terms and the plateau of the modulus of the sum of all negative terms are entirely independent of each other, and thus the possibility that the two plateaux would coincide to produce $q(\vmin)=0$ is completely unrealistic since they both depend on entirely different experimental features. Furthermore, for most realistic cases, the maximum value of $v_\text{high}^{\mu}$ is larger than the galactic escape velocity, and thus $\teta(\vmin)$ should be zero in this region. Since these plateaus cannot feasibly coincide, $q(\vmin)$ cannot equal $0$ above the largest $v_\text{high}^\mu$.

Typically isolated zeros of $q(\vmin)$ would happen when some of the ``step-like" or ``shoulder-like'' features of the modulus of the sum of negative terms in  \Eq{littleq} touch from below the monotonically increasing sum of all positive terms in \Eq{littleq}. Alternatively, if the sum of the positive terms has a region of negative curvature, it may be possible for that this sum could reach towards and touch the modulus of the sum of negative terms from above.

A practically impossible conspiracy between terms dependent on different experiments, energy intervals, and $\tilde{\eta}$ functions would be required for $q(\vmin)$ to be zero in an extended $\vmin$ interval above the minimum $v_\text{low}^{\mu}$, a conspiracy which would not survive infinitesimal changes in any of the elements defining each term in \Eq{littleq}. We include in \App{nokiss} a more mathematically rigorous proof illustrating why extended zeros of $q(\vmin)$ cannot exist above the minimum $v_\text{low}^\mu$. In the following we only consider the possibility that $q(\vmin)$ has a finite number of isolated zeros.

\subsection{Maximum number of isolated zeros of the function $q(\vmin)$ for a global likelihood }

Before counting the number of isolated zeros of $q(v_{\rm min})$, let us introduce the notion of a ``generic'' solution. We say that a solution is generic if small changes in the quantities that define it do not affect the existence of the solution. In our context, the quantities defining the solutions are the input parameters and functions given to fully specify $\xi^\text{EHI}$, $\xi^{(\alpha)}_j$, and $H_a^\text{EHI}$, \eg the efficiency function $\epsilon(E',E_R)$, the energy resolution function $G_T(\Ed,\ER)$, the differential cross section $\ud \sigma_T/ \ud E_\text{R}$, and the exposure $MT$ for each experiment and bin.

Let us briefly demonstrate the importance of the concept of generic solutions by considering the number of isolated zeros that can arise in the linear combination of two functions $f(x)$ and $g(x)$ which do not have the same functional form, since they are assumed to be derived from two independent experimental setups (\ie changes in the experimental quantities of one experiment may affect \eg $f(x)$, but do not affect $g(x)$ in the same manner). For an adjustable parameter $\lambda$, it is possible for $f(x)$ and $\lambda g(x)$ to have a generic point of osculation, \ie a point where $f(x)=\lambda g(x)$ and $f'(x)=\lambda g'(x)$, at which the Wronskian $W[f,g]$ vanishes
\begin{equation}\label{wrons1}
W[f,g](x) \equiv f(x) g'(x) - f'(x) g(x) = 0 \, .
\end{equation}
In fact, $W[f,g]$ could vanish in more than one point, say $x_1,x_2,... x_n$, or in various intervals. In this case the value of $\lambda$ can be chosen so that $f(x_1) = \lambda g(x_1)$ at one of those discrete points, say $x_1$. This point of osculation defines an isolated zero of the function $[f(x)-\lambda g(x)]$, with zero slope. Having two points of osculation, say $x_1$ and $x_2$, would require
\begin{equation} \label{wronseq}
\frac{f(x_1)}{g(x_1)} = \frac{f(x_2)}{g(x_2)} \, ,
\end{equation}
where $W[f,g](x_1)=W[f,g](x_2)=0$, for points $x_1 \neq x_2$. Small changes in the defining experimental functions and parameters would certainly break the equality in \Eq{wronseq} (or result in a non-vanishing Wronskian at those points), and thus solutions having more than one point of osculation are not generic. This same argument can be used to exclude the possibility of having a generic solution with both an isolated osculation point and an interval of osculation. Since we are interested in counting the maximum number of isolated points of osculation, we need not be concerned with the existence of intervals of osculation.

Let us denote with $X_m(\vmin)$ either the functions $H_a^\text{EHI}(\vmin)$ or the functions $\xi^{(\alpha)}_j(\vmin)$, so that \Eq{littleq} can be written in the form  
\begin{equation}
\frac{1}{2}q(\vmin) = \xi^\text{EHI}(\vmin) - \sum_{m}\lambda_m X_m(\vmin) \, ,
\end{equation}
except here we will treat the $\lambda_m$ as free parameters. The argument above ensures that there could be at most one generic point of osculation between $\xi^{\rm EHI}(v_{\rm min})$ and $\lambda_m X_m(v_{\rm min})$, or between $\lambda_m X_m(v_{\rm min})$ and $\lambda_k X_k(v_{\rm min})$ for $k\neq m$. Here, the coefficients $\lambda_m$ are adjustable parameters, equivalent to a multidimensional generalization of the parameter $\lambda$ in the above example. In the context of \Eq{littleq}, one can identify the $\lambda_m$ with the halo-dependent quantities, \eg $1/\gamma[\teta]$ and the factors in the square bracket of \Eq{Qpoisson} and \Eq{Qgauss}.

For fixed $(n-1)$ coefficients, $\lambda_1$,...$\lambda_{k-1}$, $\lambda_{k+1}$,...$\lambda_n$, we can consider two functions
\begin{equation}
f_k(\vmin) = \xi^{\rm EHI}(v_{\rm min})-\lambda_1 X_1(v_{\rm min})\cdots-\lambda_{k-1} X_{k-1}(v_{\rm min})-\lambda_{k+1} X_{k+1}(v_{\rm min})\cdots-\lambda_n X_n(v_{\rm min})
\end{equation}
and 
\begin{equation}
\lambda_k g_k(\vmin) = \lambda_k X_k(v_{\rm min}).
\end{equation}
Here we choose the parameter $\lambda_k$ with $1\leq k\leq n$ as the only adjustable parameter. Assume $\lambda_k$ can be adjusted freely. Then by adjusting $\lambda_k$, we could find one point of osculation where $f_k(\vmin) = \lambda g_k(\vmin)$, and we can treat such an adjusted value of $\lambda_k$ as a function of the rest of the parameters,
$\hat \lambda_k (\lambda_1,\dots,\lambda_{k-1},\lambda_{k+1},\dots,\lambda_n)$.

Now let us consider a $n$-dimensional manifold ${\cal M}_n$ of all the $\lambda_k$ real parameters, \ie
\begin{equation}
{\cal M}^{(n)} \equiv \{(\lambda_1,\dots,\lambda_n)| \lambda_m\in \mathbb{R} , m=1,\dots,n\}.
\end{equation}
Notice that here $n$ is one less than the total number of terms defining $q(\vmin)$ in \Eq{littleq} (because $\xi^\text{EHI}(\vmin)$ is treated separately), thus $n = \mathcal{N} = N^\text{EHI} + \sum_\alpha N^{(\alpha)}_\text{bin}$ (see \Eq{numsteps}). The equation
\begin{equation}
\lambda_k = \hat \lambda_k (\lambda_1,\dots,\lambda_{k-1},\lambda_{k+1},\dots,\lambda_n)
\end{equation}
defines a $(n-1)$-dimensional sub-manifold $M_{n-1}^{(k)}$ in the manifold ${\cal M}_n$, for each  choice of $k$.

By construction, at every point in the sub-manifold ${\cal M}_{n-1}^{(k)}$, a point of osculation
\begin{equation}
v^{(k)}(\lambda_1,\dots,\lambda_{k-1},\lambda_{k+1},\dots,\lambda_n)
\end{equation}
is assigned, so, assuming $\lambda_k$ can be adjusted freely to be $\lambda_k = \hat{\lambda}_k$, the function
\begin{eqnarray}
&f_k(\vmin) - \lambda g_k(\vmin) = \frac{1}{2} q^{(k)}(v_{\rm min}; \lambda_1,\dots,\lambda_{k-1},\lambda_{k+1},\dots,\lambda_n)\nonumber\\
 &\hspace{1.2cm}\qquad\qquad=\xi^{\rm EHI}(v_{\rm min})-\sum_{m\neq k}\lambda_mX_m(v_{\rm min})\nonumber\\
&\qquad\qquad\qquad\qquad\qquad-\hat \lambda_k (\lambda_1,\dots,\lambda_{k-1},\lambda_{k+1},\dots,\lambda_n)X_k(v_{\rm min}),
\end{eqnarray}
has at least one isolated zero (with zero slope) at
\begin{equation}
v_{\rm min}=v^{(k)}(\lambda_1,\dots,\lambda_{k-1},\lambda_{k+1},\dots,\lambda_n),
\end{equation}
for any given set of values $(\lambda_1,\dots,\lambda_{k-1},\lambda_{k+1},\dots,\lambda_n)$.

If we consider two such manifolds, ${\cal M}_{n-1}^{(k)}$ and ${\cal M}_{n-1}^{(k')}$, the intersection of them, ${\cal M}_{n-1}^{(k)}\cap{\cal M}_{n-1}^{(k')}$, is generically a $(n-2)$-dimensional sub-manifold. Assuming now that $\lambda_k$ and $\lambda_{k^\prime}$ ($k\neq k^\prime$) can both be adjusted at will so that $\lambda_k = \hat\lambda_k$ and $\lambda_{k^\prime} = \hat\lambda_{k^\prime}$ at every point in this sub-manifold, we have two isolated zeros (with zero slope) given by the functions,
\begin{eqnarray}
&v^{(k)}(\lambda_1,\dots,\lambda_{k-1},\lambda_{k+1},\dots,\lambda_{k'-1},\lambda_{k'+1},\dots,\lambda_n)\nonumber\\
&\qquad\equiv v^{(k)}(\lambda_1,\dots,\lambda_{k-1},\lambda_{k+1},\dots,\lambda_n)|_{\lambda_{k'}=\hat\lambda_{k'}},
\end{eqnarray}
and
\begin{eqnarray}
&v^{(k')}(\lambda_1,\dots,\lambda_{k-1},\lambda_{k+1},\dots,\lambda_{k'-1},\lambda_{k'+1},\dots,\lambda_n)\nonumber\\
&\qquad\equiv v^{(k')}(\lambda_1,\dots,\lambda_{k'-1},\lambda_{k'+1},\dots,\lambda_n)|_{\lambda_{k}=\hat\lambda_{k}},
\end{eqnarray}
which are respectively induced from the functions defined on ${\cal M}_{n-1}^{(k)}$ and ${\cal M}_{n-1}^{(k')}$. The values of these two functions at the same point are in general different.

In a similar way, if all coefficients $\lambda_k$ could be freely adjusted the intersection of all $(n-1)$-dimensional sub-manifolds,
\begin{equation}
\cap_{k=1}^n{\cal M}_{n-1}^{(k)} \, ,
\end{equation}
is generically a zero-dimensional sub-manifold of ${\cal M}_{n}$, i.e. a set of discrete points. For one of these points, which we call $(\hat \lambda_1,\dots,\hat \lambda_n)$, we can define the function
\begin{equation}\label{qproof}
\frac{1}{2}q(v_{\rm min};\hat \lambda_1,\dots,\hat \lambda_n) = \xi^{\rm EHI}(v_{\rm min})-\sum_{m=1}^n
\hat\lambda_m X_m(v_{\rm min})
\end{equation}
which has $n$ isolated zeros, with zero slope. Here, $n=\mathcal{N}\equiv N_\text{EHI} +\sum_\alpha N^{(\alpha)}_{\rm bin}$ (see \Eq{numsteps}), \ie the number of events observed by the EHI experiment plus the total number of bins employed by all Poisson and Gaussian experiments. This is what we wanted to prove. However we have so far assumed the coefficients $\lambda_m$ could all be freely adjusted. This is not true, however, and the actual number of isolated zeros of $q(\vmin)$ (with $q'(\vmin)=0$) will be in most circumstances much smaller than the maximum $\mathcal{N}$.

In fact, the coefficients $\lambda_m$ are quantities derived from a halo function $\tilde\eta$. All points in ${\cal M}_n$ that can be actually realized from halo functions $\tilde\eta$ form a continuous subset ${\cal S}$ of the manifold ${\cal M}_n$. The maximum number of the individual sub-manifolds ${\cal M}_{n-1}^k$ passing through a point in ${\cal S}$ gives the maximum number of actual possible steps in the best fit $\tilde\eta$ function. This number can be determined by carefully considering the functional form of $\xi^\text{EHI}$, $\xi_j^{(\alpha)}$, and $H_a^{\rm EHI}$, and is in general smaller than $\mathcal{N}$.

\subsection{Argument against non-isolated zeros of $q(\vmin)$\label{nokiss}}
Here, we provide a more mathematically rigorous proof for why $q(\vmin)$ cannot have non-isolated zeros above the minimum $v_\text{low}^\mu$. 

Using \Eq{eq:qvmin1} we can equate the functional derivative of $L[\teta]$ to the derivative of $q(\vmin)$ as 
\begin{equation}
\frac{\delta L}{\delta \teta(\vmin)} = \frac{\partial}{\partial \vmin}q(\vmin) \, . 
\end{equation}
We will begin by assuming that there exists some interval $[v_1,v_2]$ above the minimum $v_\text{low}^\mu$ in which $q(\vmin)=0$, and prove by contradiction that this cannot be the case. 

Let us introduce an infinitesimal perturbation $\delta F(v)$ in the speed distribution $F(v)$, that is only non-zero in the interval $[v_1,v_2]$, and define the quantity 
\begin{equation}\label{Delta}
\Delta \equiv \frac{\rho \sigma_\text{ref}}{m}\int_{v_\delta}^{\infty} \ud v \frac{\delta F(v)}{v}q(v) \, .
\end{equation}
Since $q(v) =0$ in the interval $[v_1,v_2]$ and $\delta F(v)=0$ outside of this interval, $\Delta$ is trivially zero. Since the halo function $\teta$ linearly depends on $F(v)$, the induced change in the halo function $\teta$ is 
\begin{equation}
\delta \teta(\vmin) \equiv \frac{\rho \sigma_\text{ref}}{m}\int_{\vmin}^{\infty}\ud v \frac{\delta F(v)}{v} \, ,
\end{equation}
and its derivative is given by
\begin{equation}
\frac{\partial}{\partial v}\delta \teta(v) = - \frac{\rho \sigma_\text{ref}}{m}\frac{\delta F(v)}{v} \, .
\end{equation}
Performing integration-by-parts on the integral in \Eq{Delta} gives 
\begin{eqnarray}
\Delta & = & \int_{v_\delta}^{\infty} \ud v \left[-\frac{\partial}{\partial v}\delta \teta\right]q(v) \\
& = & -\delta\teta(\infty)q(\infty) + \delta\teta(v_\delta)q(v_\delta) + \int_{v_\delta}^{\infty}\ud v \delta \teta \left[ \frac{\partial}{\partial v}q(v)\right] \\
& = & \int_{v_\delta}^{\infty} \ud v \delta \teta \frac{\delta L}{\delta \teta(v)} \\
& = & L[\teta + \delta \teta] - L[\teta] \equiv \delta L \, ,
\end{eqnarray}
where the last line is obtained from the definition of the functional derivative.

This expression for $\Delta$ implies that any change of $F(v)$ introduced above the
minimum $v_\text{low}^\mu$ in an interval where $q(\vmin) = 0$ would not change the value of the likelihood. However, by definition, this perturbation necessarily introduces a constant shift in $\teta$ at all $\vmin$ values below $v_1$. It is inconceivable for a likelihood function to be invariant under such a rigid shift of $\teta$. Therefore, our original assumption must have been false, and there cannot exist intervals above the minimum $v_\text{low}^\mu$ in which $q(\vmin)=0$.

The interested reader might wonder why the proof presented above does not preclude the existence of isolated zeros. In this case, the modification to the speed distribution $\delta F(v)$ must take the form of a constant times a delta function (\ie nonzero only at the location of the isolated zero). The resultant speed distribution $F(v)+\delta F(v)$ would no longer be a smooth function, and thus cannot possibly be representative of a true speed distribution (no physical process would produce a component with zero dispersion).

\section{The uniqueness of the best-fit halo function}
Here, we show that the halo function  $\tilde\eta_{BF}(v_{\rm min})$ maximizing a global likelihood functional having at least one extended likelihood as a factor is unique in the $v_{\rm min}$ range wherein the experiments in consideration can probe the value of the halo function. The proof consists in showing that the second directional derivatives  of  the functional $L\equiv -2\ln{\cal L}$ with respect to variations of the $\tilde\eta$ function are all positive.

\subsection{Statement of the proof}

We start by stating two properties of the global likelihood (and the individual likelihoods) considered in this paper. First, the likelihood depends on the halo function only through physically observable quantities, which are either the scattering rate in a bin
\begin{equation}
R_j^{(\alpha)}\equiv MT\int_{v_\delta}^\infty{\rm d}v~{\cal R}_{[E'_j,E'_{j+1}]}(v)\tilde\eta(v),
\end{equation}
or the value of the differential rate at a given value of $E'$,
\begin{equation}
\frac{{\rm d}R}{{\rm d}E'}\equiv MT\int_{v_\delta}^\infty{\rm d}v~\frac{{\rm d}{\cal R}}{{\rm d}E'}(v)\tilde\eta(v) \, .
\end{equation}
Secondly, if we treat these observable quantities (which we call ``the rates" in the rest of this section) as independent parameters without restriction, the global likelihood ${\cal L}$ is a strictly concave function of them, or, equivalently, the functional $L\equiv -2\ln{\cal L}$ is a strictly convex function.

Since the rates depend linearly on the halo function, the functional $L$ is a convex but not necessarily a strictly convex function of the halo function $\tilde\eta$. The strict convexity of the functional $L$ as a function of the rates guarantees the uniqueness of the best fit rates ({\bf{those}} which maximize the likelihood ${\cal L}$, and thus minimize the functional $L$), but not of the best fit halo function $\tilde\eta_{BF}(v_{\rm min})$, since, in general, the same values of the rates can be obtained from different halo functions.

While these two properties do not yet prove that the best fit halo function is unique, we know from the convexity of the functional $L$ that if there are more than one best fit halo functions, the value of the likelihood is constant along the line of minima between any two best fit halo functions, and thus, along the direction of the line, the second (and also higher) order directional (functional) derivatives should vanish. Thus
all the best fit halo functions are connected to each other by a continuous deformation and thus form a connected set. 

Using this fact, the global uniqueness of the best fit halo function $\tilde\eta_{BF}(v_{\rm min})$ can be asserted by proving that  the second order  directional derivatives  of $L$ around a minimum are all larger than zero, i.e. 
\begin{equation} \label{hessian}
\int{\rm d}v\int{\rm d}w~\Delta\tilde\eta(v)\Delta\tilde\eta(w)\left.\frac{\delta^2}{\delta\tilde\eta(v)\delta\tilde\eta(w)} L[\tilde\eta]\right|_{\tilde\eta=\tilde\eta_{\rm BF}}>0,
\end{equation}
for all allowed $\Delta\tilde\eta$. 

Up to this point, we have not used the fact that the halo function is a non-increasing function, and that the KKT conditions should be satisfied by the best fit halo function.
Since we have previously proven in Appendix A that the halo functions maximizing the global likelihood are piecewise constant with at most ${\cal N}=N_\text{EHI}+\sum_{\alpha}N_{\rm bins}^{(\alpha)}$ points of discontinuity, we know that deformations of $\tilde\eta$ between two best fit halo functions must also respect this form. Thus, the positivity condition of the second order  directional derivatives  of $L$ around a minimum, Eq.~\ref{hessian}, can be rewritten as
\begin{eqnarray} \label{hessian2}
0<&\sum_{a,b=1}^{\cal N}
 \Big[   
\Delta\tilde\eta_a\Delta\tilde\eta_b
\frac{\delta^2}{\partial\tilde\eta_a\partial\tilde\eta_b}
f_{L}(\{\vec v,\vec{\tilde\eta}\})\nonumber\\
&+
\Delta v_a\Delta v_b
\frac{\delta^2}{\partial v_a\partial v_b}
f_{L}(\{\vec v,\vec{\tilde\eta}\})\nonumber\\
&+
2\Delta\tilde\eta_a \Delta v_b
\frac{\delta^2}{\partial\tilde\eta_a\partial v_b}
f_{L}(\{\vec v,\vec{\tilde\eta}\})
\Big]_{\tilde\eta=\tilde\eta_{\rm BF}},
\end{eqnarray}
for all allowed infinitesimal variations $\Delta \tilde\eta_a$ and $\Delta v_a$, where $f_{L}(\{\vec v,\vec{\tilde\eta}\})\equiv{L}[\tilde\eta(v_{\rm min};\{\vec v,\vec{\tilde\eta}\})]$.

When one finds the best fit halo function, the locations $v_a$ and heights $\tilde\eta_a$ of the steps can be independently varied since the KKT conditions are automatically satisfied for the resultant best fit halo function. However, in Eq.~\ref{hessian2}, if a variation of these parameters, $\Delta\tilde\eta_a$ and $\Delta v_a$, truly connects two best fit halo functions (and the ones between them, which are also best fit halo functions) by a continuous deformation, the KKT conditions should remain satisfied along the path of the deformation, and thus we only need to consider the variations respecting the KKT conditions.
We will show now that these variations must all have  $\Delta v_a=0$, namely the positions of the steps in $v_{\rm min}$ cannot change.

\subsection{Proof that  the locations of the steps cannot change}

Let us examine how the function $q(v)$ changes under an arbitrary deformation $\Delta\tilde\eta(v_{\rm min})$ of the halo function $\tilde\eta(v_{\rm min})$.
Using the definition of the function $q(v)$ in Eq.~(3.6), the induced variation  $\Delta q(v)$ of the function $q(v)$, can be compactly written in terms of the second derivative of the functional $L$ as
\begin{eqnarray}
\Delta q(v)&\equiv& \int_{v_\delta}^\infty{\rm d}w~\Delta\tilde\eta(w)\frac{\delta q(v)}{\delta \tilde\eta(w)}\nonumber\\
&=& \int_{v_\delta}^\infty{\rm d}w~\Delta\tilde\eta(w)\frac{\delta}{\delta \tilde\eta(w)}
\left[\int_0^v{\rm d}u\frac{\delta L}{\delta\tilde\eta(u)}\right],
\end{eqnarray}
or
\begin{eqnarray}
\label{q-variation}
\Delta q(v)&=& \sum_{\alpha,j}\Delta R^{(\alpha)}_j\int_{v_\delta}^v{\rm d}u\frac{\delta}{\delta\tilde\eta(u)}\left(\frac{\partial L}{\partial R^{(\alpha)}_j}\right)\nonumber\\
&&+\int{\rm d}E'~\Delta \left(\frac{{\rm d}R}{{\rm d}E'}\right)\int_{v_\delta}^v{\rm d}u\frac{\delta}{\delta\tilde\eta(u)}\left(\frac{\partial L}{\partial ({\rm d}R/{\rm d}E')}\right)
\end{eqnarray}
where we have defined the  changes $\Delta R_j^{(\alpha)}$ and  $\Delta \left(\frac{{\rm d}R}{{\rm d}E'}\right)$ of the rate $R_j^{(\alpha)}$ and differential rate ${{\rm d}R}/{{\rm d}E'}$,  respectively as
\begin{equation}
\Delta R_j^{(\alpha)}\equiv\int{\rm d}v~\Delta\tilde\eta(v)\frac{\delta R_j^{(\alpha)}}{\delta \tilde\eta(v)}
\end{equation}
and 
\begin{equation}
\Delta \left(\frac{{\rm d}R}{{\rm d}E'}\right)\equiv\int{\rm d}v~\Delta\tilde\eta(v)\frac{\delta  }{\delta \tilde\eta(v)}\left(\frac{{\rm d}R}{{\rm d}E'}\right).
\end{equation}
Eq.~(\ref{q-variation}) shows that the function $q(v)$ is invariant under a variation $\Delta\tilde\eta$ of the best fit halo function that leaves the rates unchanged (all best fit halo functions should yield the same unique best fit rates). This implies that all best fit halo functions must have their points of discontinuity (i.e. the locations of its steps) at the same $v_{\rm min}$ values. In other words, a variation having a non-zero 
$\Delta v_a$ either breaks the KKT condition or changes the observable rates, and thus such a variation inevitably decreases the value of the likelihood functional.

\subsection{Evaluation of the second directional derivatives of $L$ }

Since the positions of the steps cannot change, it is enough to evaluate the second derivative of the functional $L$ with respect to an arbitrary variation of the heights of the steps $\Delta\tilde\eta_a$. 
  Expanding the functional $L$ around the best fit halo up to the second order we get
\begin{eqnarray}
 \label{L-variation}
\Delta L&=&\frac{1}{2}\sum_{i,j=1}^{\cal N}\Delta\tilde\eta_i\Delta\tilde\eta_j\int_{v_{i-1}}^{v_i}{\rm d}v~\int_{v_{j-1}}^{v_j}{\rm d}u\frac{\delta^2 L}{\delta\tilde\eta(v)\delta\tilde\eta(u)}\\
&=&\frac{1}{2}\sum_{i,j=1}^{\cal N}\Delta\tilde\eta_i\Delta\tilde\eta_j\int_{v_{i-1}}^{v_i}{\rm d}v~\frac{\delta}{\delta\tilde\eta(v)}\left(q(v_j)-q(v_{j-1})\right)\\
&=&2\sum_{A=1}^{\cal N}\left(\sum_{i=1}^{\cal N} {\cal K}_{Ai}\Delta\tilde\eta_i\right)^2.
\end{eqnarray}
Here we defined the index $A$ to run over all data points, namely all bins or single events of all experiments considered, so $A$ runs from 1 to ${\cal N}$. Specifically,
the summation over $A$ runs over the observed  events in the extended likelihoods and the bins $j$ of all experiments $\alpha$ with Poisson and Gaussian likelihoods.  The index $i$ indicates instead each constant portion of the best fit halo function, between the steps at $v_{\rm min}$ values $v_{i-1}$ and $v_i$.
The maximum number of steps was found in Appendix A to be ${\cal N}$, so we can take take the number of steps  to be equal to   ${\cal N}$  and  consider some of the step heights to be zero. In this way, 
 $i$ also runs from 1 to ${\cal N}$.   The coefficients $K_{Ai}$ are given by 
\begin{equation}
{\cal K}_{Ai}\equiv \, \frac{H_A(v_i)-H_A(v_{i-1})}{\gamma_A[\tilde\eta]}
\end{equation}
for extended likelihoods,
\begin{equation}
{\cal K}_{Ai}\equiv \sqrt{ n_j^{(\alpha)}} \, \frac{\xi^{(\alpha)}_j(v_i)-\xi^{(\alpha)}_j(v_{i-1})}{\nu^{(\alpha)}_j[\tilde\eta]+b^{(\alpha)}_j}
\end{equation}
for Poisson likelihoods, and
\begin{equation}
{\cal K}_{Ai}\equiv \, \frac{\xi^{(\alpha)}_j(v_i)-\xi^{(\alpha)}_j(v_{i-1})}{\sigma^{(\alpha)}_j} \, ,
\end{equation}
for Gaussian likelihoods. In the last two equations the index $A$ accounts for the experiment-bin pairs indexes $(\alpha, j)$.

Notice that the quantities ${\cal K}_{Ai}$ can be interpreted as the components of ${\cal N}$  non-zero vectors $\vec{\cal K}_{A}$  in a vector space with dimension ${\cal N}$  with components denoted by $i$. We can also consider $\Delta\tilde\eta_i$ to be the components of a vector $\Delta\vec{\tilde\eta}$ with the same number of dimensions  of the $\vec{\cal K}_{A}$ vectors. Each vector $\Delta\vec{\tilde\eta}$ is a possible infinitesimal variation of the heights of the steps $\tilde\eta_i$ around a best fit halo function. Eq.~(\ref{L-variation}) is then a sum of the squares of the inner products of two vectors $\vec{\cal K}_{A}$ and $\Delta\vec{\tilde\eta}$.

Notice also that the vectors $\vec{\cal K}_{A}$ are generically linearly independent, because there is no reason that the experiment-specific quantities ${\cal K}_{Ai}$ should be dependent upon information contained in a different bin or experiment.
 
  Since the vectors $\vec{\cal K}_{A}$ are generically linearly independent there is no non-zero vector $\Delta\vec{\tilde\eta}$ orthogonal to all of them. This implies that there is no infinitesimal variation of the heights of the steps $\tilde\eta_i$ around a best fit halo function for which the second order variation of $L$ vanishes.  This proves that the likelihood functional $L$ is not invariant under any infinitesimal variation around the best fit halo function which would lead to another best fit halo function. Since all the second directional derivatives of $L$ around a best fit halo function are positive the best fit halo function must be unique.

\bibliographystyle{JHEP}
\bibliography{biblio}

\end{document}